\documentclass[twocolumn]{aastex631}

\usepackage{natbib}
\usepackage{graphicx}
\usepackage{hyperref}
\usepackage{amsmath}
\usepackage{lipsum}
\usepackage{grffile}
\usepackage{hyphenat}
\usepackage{longtable} 

\begin{document}

\title{The Absolute Age of Milky Way Globular Clusters}

\author{Jiaqi (Martin) Ying}
\affiliation{Department of Physics and Astronomy, Dartmouth College, 6127 Wilder Laboratory, Hanover, NH 03755, USA}

\author{Brian Chaboyer} 
\affiliation{Department of Physics and Astronomy, Dartmouth College, 6127 Wilder Laboratory, Hanover, NH 03755, USA}

\author{Michael Boylan-Kolchin}
\affiliation{Department of Astronomy, The University of Texas at Austin, TX 78712, USA}

\author{Daniel R. Weisz}
\affiliation{Department of Astronomy, University of California Berkeley, CA 94720, USA}

\author{Rowan Goebel-Bain}
\affiliation{Department of Physics and Astronomy, Dartmouth College, 6127 Wilder Laboratory, Hanover, NH 03755, USA}

\begin{abstract}
    Globular clusters (GCs) provide statistically significant coeval populations of stars spanning various evolutionary stages, allowing robust constraints on stellar evolution model parameters and ages. We analyze eight old Milky Way GCs with metallicities between [Fe/H] $=-2.31$ and $-0.77$ by comparing theoretical isochrone sets from the Dartmouth Stellar Evolution Program to HST observations. The theoretical isochrones include uncertainties introduced by $21$ stellar evolution parameters such as convective mixing, opacity, diffusion, and nuclear reactions, capturing much of the quantifiable physics used in our code. For each isochrone, we construct synthetic color-magnitude diagrams (CMD) near the main-sequence turn-off region and apply two full-CMD-fitting methods to fit HST ACS data across a range of distance and reddening and measure the absolute age of each GC from the resulting posterior distribution, which accounts for uncertainties in the stellar models, observations, and fitting method. The resulting best-fitting absolute ages range from $\approx 11.5$ to $13.5$ Gyr, with a typical error of $0.5-0.75$ Gyr; the data show a clear trend toward older ages at lower metallicities. Notably, distance and reddening account for over $50\%$ of the uncertainty in age determination in each case, with metallicity, $\alpha$ abundance, mixing length, and helium diffusion being the most important stellar physics parameters for the error budget. We also provide an absolute age-metallicity relation for Milky Way GCs.
\end{abstract}

\section{Introduction}
Globular clusters (GCs) are gravitationally bound clusters of stars and are among the oldest objects in the universe. Their relatively stable structure, co-evolutionary nature, and high stellar density make them ideal astronomical laboratories for addressing critical problems in galaxy formation, stellar evolution, and cosmology. GCs contain large numbers of stars at various evolutionary stages, providing statistically significant samples to calibrate stellar evolution models \cite[e.g.][]{dotterDartmouthStellarEvolution2008,vandenbergAges55Globular2013}. Furthermore, GCs also host some of the oldest black holes \cite[e.g.][]{giesersDetachedStellarmassBlack2018,giesersStellarCensusGlobular2019}, and dark central masses, which could be intermediate-mass black holes \cite[e.g.][]{vitralElusiveDarkCentral2023}. The well-calibrated properties of GCs, such as age, stellar mass, and kinematics \citep{haberleFastmovingStarsIntermediatemass2024}, help to quantify the properties of black holes, which, for example, can be used to place strong constraints on the cosmological coupling between black holes, and an expanding universe \citep{rodriguezConstraintsCosmologicalCoupling2023}.

The large sample size of stars also makes GCs an essential and reliable observational foundation for understanding composite stellar populations both inside and outside of the Milky Way, which serves as a building block for the stellar population synthesis \citep{bicaBaseStarClusters1986}. The age-metallicity relation of GCs is critical for understanding the formation and evolutionary history of galaxies, as well as the processes of chemical enrichment in the universe \cite[e.g.][]{marin-franchACSSurveyGalactic2009,dotterACSSurveyGalactic2010}. The bimodality of GCs (each galaxy generally has a metal-poor and a metal-rich sub-population) provides insight into galaxy evolution history \citep{arnoldFossilRecordTwophase2011} and GC formation \citep{el-badryFormationHierarchicalAssembly2019}. 

Using JWST data, \citet{mowlaSparklerEvolvedHighredshift2022} found proto-GCs formed at $z > 9$ ($\sim 0.5$ Gyr after the big bang) and \citet{adamoDiscoveryBoundStar2024} found young massive star clusters at $z \sim 10.2$ ($\sim 0.46$ Gyr after the big bang). Despite having an uncertainty of up to 10\% in the age estimate derived from redshift \citep{boylan-kolchinUncertainTimesRedshifttime2021}, these discoveries are consistent with the fact that GCs are the oldest objects whose ages may be accurately determined and can be used to set the lower limit of the age of the universe independent of the cosmological model \cite[e.g.][]{chaboyerAccurateRelativeAge1996,kraussAgeEstimatesGlobular2003}.

Accurately determining the ages of GCs is central to many scientific endeavors. Forward modeling based on stellar evolution code and isochrone fitting with observational data on the color-magnitude diagrams (CMDs) remains the most robust age estimation method. Most studies rely on pre-existing isochrone databases such as Dartmouth Stellar Evolution Database \citep{dotterDartmouthStellarEvolution2008}, MESA Isochrones \& Stellar Tracks \citep{dotterMESAISOCHRONESStelLAR2016, choiMesaIsochronesStellar2016}, etc., because of the large amount of computational resources and sophisticated knowledge required to generate a large grid of stellar models and isochrones. As a result, those studies rely on a single set of stellar evolution parameters (e.g., nuclear reaction rates, opacity, treatment of convection) despite inherent, and sometimes substantial, uncertainties in these parameters (see Table.~2 from \citealt{yingAbsoluteAgeNGC2024}). As a result, most GC age studies focus on relative ages \cite[e.g.][]{stetsonRelativeAgesGalactic1996, marin-franchACSSurveyGalactic2009}. Some estimates suggest that the ``zeropoint'' uncertainties in the reported absolute ages of GCs are $\sim 1.5\,$Gyr \citep[e.g.][]{chaboyerTestingMetalPoorStellar2017,omalleyAbsoluteAgesDistances2017}, which is equivalent to being unable to distinguish between the redshifts of $z = 3$ and $z = 8$ (i.e., the difference between a GC forming before or after cosmic reionization). Because underlying stellar physics varies with factors such as chemical composition \citep[e.g.][]{bonacaCalibratingConvectiveProperties2012, tannerEffectMetallicitydependentTensuremathtau2014}, estimating and correcting for such age inaccuracies is non-trivial and must be done on a cluster-by-cluster basis.

Absolute ages of GCs are scientifically valuable as researchers can use them to constrain the age of the universe and study the early formation history of our galaxy. In contrast to relative ages, absolute ages must account for intrinsic uncertainties in the stellar models and extrinsic uncertainties in distance, reddening, and composition measurements. Achieving this has been a longstanding goal \citep{chaboyerAccurateRelativeAge1996}. Recent advances in computational power and the increasing availability of observational constraints provide an opportunity to estimate the absolute ages of GCs with improved precision.

In this paper, we estimate the absolute age of eight Milky Way GCs: NGC  104 (47 Tuc), 4147, 5053, 5466, 6362, 6809 (M55), 7078 (M15), and 7099 (M30). In \S \ref{observational_data}, we introduce the observational data used to calibrate or fit isochrones. In \S \ref{iso_construct}, we explain the process of evolving models and constructing isochrones. In \S \ref{iso_fitting}, we review the historical isochrone-fitting methods and introduce the full-CMD-fitting methods used in this paper. In \S \ref{results}, we present our main results and analyze intrinsic and extrinsic uncertainties.

\section{Observational data} \label{observational_data}

\begin{table*}[!htbp]\centering
\caption{Universal Monte Carlo Input Parameters \label{tab_uni_MC}}
\begin{tabular}{llll}
\hline
Variable                               & Distribution & Range        & Source                                               \\
\hline
$\Delta Y/\Delta Z$ & Uniform & $1.75 \sim 2.5$ & \citet{peimbertPrimordialHeliumAbundance2016}\\
Helium abundance   & Uniform      & $0.2465(25) + \left(\Delta Y/\Delta Z\right) Z$   & \citet{averEffectsHeTextbackslashuplambda108302015}                                       \\
Mixing length                 & Uniform      & $1.0 \sim 2.5$      & N/A                                         \\
Heavy element diffusion       & Uniform      & $0.5 \sim 1.3$      & \citet{thoulElementDiffusionSolar1994}                                         \\
Helium diffusion              & Uniform      & $0.5 \sim 1.3$    & \citet{thoulElementDiffusionSolar1994}                                           \\
Surface boundary condition &   Trinary          &     1/3; 1/3; 1/3             & \citet{eddingtonInternalConstitutionStars1926}\\
  &             &                     & \citet{krishnaswamyProfilesStrongLines1966} \\
  &             &                     & \citet{hauschildtNextGenModelAtmosphere1999} \\
Low temperature opacities     & Uniform      & $0.7 \sim 1.3$        & \citet{fergusonLowTemperatureOpacities2005}                                      \\
High temperature opacities    & Normal       & $1.0 \pm 0.03$           & \citet{iglesiasUpdatedOpalOpacities1996}     \\
Plasma neutrino loses         & Normal       & $1.0 \pm 0.05$     & \citet{haftStandardNonstandardPlasma1994}     \\
Conductive opacities          & Normal       & $1.0 \pm 0.20$       & \citet{hubbardThermalConductionElectrons1969}\\
 &             &                     & \citet{canutoElectricalConductivityConductive1970}   \\
Convective envelope overshoot & Uniform      & $0 \sim 0.2$      & N/A                                          \\
Convective core overshoot & Uniform & $0 \sim 0.2$ & N/A \\
$p + p \to H_2 + e + \nu$   
& Normal       & $\left(4.07 \pm 0.04 \right)\times 10^{-22}$ & \citet{acharyaUncertaintyQuantificationProtonproton2016}\\
 & & & \citet{marcucciProtonProtonWeakCapture2013}\\
${ }^{3}He + { }^{3}He \to { }^{4}He + p + p$                 & Normal       & $5150 \pm 500$& \citet{adelbergerSolarFusionCross2011}\\
${ }^{3}He + { }^{4}He \to { }^{2}H + \gamma$                  & Normal       & $0.54 \pm 0.03$&\citet{deboerMonteCarloUncertainty2014}\\
${ }^{12}C + p \to { }^{13}N + \gamma$                & Normal       & $1.45 \pm 0.50$ & \citet{xuNACREIIUpdate2013}\\
${ }^{13}C + p \to { }^{14}N + \gamma$              & Normal       & $5.50 \pm 1.20$& \citet{chakrabortySystematicRmatrixAnalysis2015}\\
${ }^{14}N + p \to { }^{15}O + \gamma$             & Normal       & $3.32 \pm 0.11$ & \citet{martaN14pensuremathgammaO15ReactionStudied2011}\\
${ }^{16}N + p \to { }^{17}F + \gamma$               & Normal       & $9.40 \pm 0.80$ & \citet{adelbergerSolarFusionCross2011}\\
\hline
\end{tabular}
\end{table*}

\subsection{Calibration Stars}

\begin{table}[!htbp]
\caption{Calibration Stars \label{tab_cali_star}}
\resizebox{\columnwidth}{!}{\begin{tabular}{l|llll}
\hline
Clusters & Single Star                               & F606W        & F606W - F814W    & [Fe/H]                               \\
\hline
NGC 6809 (M55) & HD103269                          & $5.809(4)$     &  $0.581(4)$  & -1.83\\   
NGC 4147 & HD108200                         & $6.249(5)$     &  $0.663(6)$ &-1.83 \\ 
\hline
NGC 7078 (M15) & HD46120                           & $5.787(3)$     &  $0.566(2)$ & -2.22\\ 
NGC 7099 (M30) & HD106924                          & $6.041(4)$     &  $0.601(5)$ & -2.23\\
NGC 5053  & & \\
NGC 5466 & & \\
\hline
\end{tabular}}
\end{table}

Distance is one of the biggest sources of uncertainty when determining the absolute age of GCs. \citet{yingAbsoluteAgeM922023} shows that the uncertainty of distance to M92 contributes $80\%$ to the total age error. \textit{Gaia} Early Data Release 3 (EDR3) from the \textit{Gaia} mission \citep{gaiacollaborationGaiaEarlyData2021} provides parallax information with reduced uncertainty \citep{vasilievGaiaEDR3View2021}. However, most Milky Way GCs are located in the outer regions of the galaxy, presenting observational challenges for \textit{Gaia} mission. As a result, the distance uncertainties derived from \textit{Gaia} do not offer significant advantages over other methods, such as HST kinematic distances, subdwarf distances, star count distances, or RR Lyrae distances \citep{baumgardtAccurateDistancesGalactic2021}.

Metal-poor main-sequence stars with well-determined distances can be used to test and calibrate the stellar isochrones in an age-independent manner. For this, we use calibration stars, as shown in Table~\ref{tab_cali_star}. These calibration stars have precise HST Advanced Camera for Surveys (ACS) photometry and compositions that closely match those of the GCs being calibrated. The selected stars are main-sequence stars and are not members of multiple star systems, making them ideal targets for calibrating the physics in stellar evolution models. The HST photometry is presented in \citet{chaboyerTestingMetalPoorStellar2017} and the high-resolution spectroscopic abundances in \citet{omalleyDifferentialAbundanceAnalysis2017}. Those stars have accurate \textit{Gaia} EDR3 parallaxes. The uncertainty in \textit{Gaia} EDR3 parallax can be converted to the uncertainty in the distance modulus as:
\begin{equation}
    \delta_{\mu} \approx 5 \frac{\delta_p}{p \ln{(10)}} \label{eq1},
\end{equation}
where $\mu$ is distance modulus and $p$ is \textit{Gaia} EDR3 parallax. Because those calibration stars are located much closer to us than the GCs being calibrated, they have much smaller uncertainty in distance and, therefore, absolute luminosity. The EDR3 parallaxes are known to suffer from systematic zero-point errors, and a calibration of this error has been given by   \citet{lindegrenGaiaEarlyData2021a}. However, the zero-point correction is not that well calibrated for bright stars like these two stars \citep{lindegrenGaiaEarlyData2021a}, and there is evidence that the zero-point correction may be an over-correction for bright stars \citep{riessCosmicDistancesCalibrated2021,zinnValidationGaiaEarly2021} so we elected to add in half the zero-point correction to the quoted EDR3 parallax. The uncertainty in the parallaxes was taken as the value of the zero-point correction added in quadrature, and the uncertainty in the parallax was given in EDR3. 

\subsection{Photometric Data} \label{HST_data}
We utilize calibrated photometric data from the Hubble Space Telescope (HST) Advanced Camera for Surveys (ACS) globular cluster survey treasury program \citep{sarajediniACSSurveyGalactic2007,andersonACSSURVEYGLOBULAR2008}.
The ACS GC survey included artificial star tests that estimate the photometric uncertainties and completeness as a function of magnitude and cluster position \citep{andersonACSSURVEYGLOBULAR2008}. 
We use a subset of stars around the main sequence turn-off (MSTO) whose position is most sensitive to variations in age, and relatively insensitive to the present-day mass function \citep{chaboyerAccurateRelativeAge1996} to compare with theoretical models.. Specifically, we select stars within $\pm 2\,$magnitudes from the point on the subgiant branch, which is $0.05\,$mag redder than MSTO. Additionally, we remove blue straggler stars and outliers by selecting stars within $0.08\,$mag in F606W of the median ridgeline in a magnitude-magnitude diagram of F814W and F606W. We choose $8$ GCs with low reddening to avoid the effect of differential reddening, and \citet{legnardiDifferentialReddeningDirection2023} confirms that all $8$ GCs in this study do not suffer from differential reddening.

\subsection{Detach Eclipsing Binaries}

\begin{table}[!htbp]
\caption{Detach Eclipsing Binaries \label{tab_DEBs}}
\resizebox{\columnwidth}{!}{\begin{tabular}{l|llll}
\hline
Clusters & DEBs & Mass  & Luminosity     & Radius                                     \\
\hline
NGC 6809 (M55) & V54p & 0.726(15) & 1.381(67) & 1.006(9) \\
& V54s & 0.555(8) & 0.159(12) & 0.528(5) \\
\hline
NGC 104 (47 Tuc) & E32p & 0.8617(47) & 1.65(5) & 1.1834(34) \\
               & E32s & 0.8268(45) & 1.14(4) & 1.0045(40) \\
               & V69p & 0.8762(48) & 1.96(6) & 1.3148(51) \\
               & V69s & 0.8588(60) & 1.56(5) & 1.1616(62) \\
\hline
NGC 6362 & V40p & 0.8337(63) & 2.27(18)  & 1.3253(77) \\
        & V40s & 0.7947(48) & 1.24(13)  & 0.997(13)  \\
        & V41p & 0.8215(58) & 1.46(11)  & 1.0739(48) \\
        & V41s & 0.728(47)  & 0.524(45) & 0.7307(46)\\
\hline
\end{tabular}}
\end{table}

Since calibration stars are not members of the GC calibrated, they can only be used to calibrate the physics in stellar evolution models rather than to infer GC parameters directly. Some GCs, however, also host detached eclipsing binaries (DEBs). DEB analysis utilizes photometric and spectroscopic data as well as light and radial velocity curves to determine the fundamental stellar parameters (masses, luminosities, and radii) of the components of GC binaries to a precision better than $1 \%$ \citep{kaluznyEclipsingBinariesGlobular2005}. More importantly, DEB analysis does not make assumptions on the distance, and DEBs can be modeled as single stars, which can be used to fit isochrones as an independent age estimation method \citep{yingAbsoluteAgeNGC2024}.  Few GCs have DEBs with known properties, and even fewer GCs have DEBs in the age-sensitive phase of evolution (main-sequence turn-off and giant branch). Table~\ref{tab_DEBs} lists the masses, luminosities, and radii of DEBs derived from (\citet{kaluznyClustersAgeSExperiment2014}, M55), (\citet{thompsonClusterAgeSExperiment2020}, 47 Tuc), and (\citet{kaluznyClusterAgesExperiment2015}, NGC  6362) used in this study.

\section{Isochrone Construction} \label{iso_construct}

\begin{table*}[!htbp]
\caption{Individual Monte Carlo Input Parameters with Distance and Reddening\label{tab_indi_MC}}
\centering
\resizebox{0.7\textwidth}{!}{%
\begin{tabular}{lllll}
\hline
Cluster & Variable                               & Distribution & Range        & Source                                               \\
\hline
NGC 6809 (M55) & {[}Fe/H{]}                    & Normal       & $-1.90 \pm 0.10$     &  \cite{rainChemicalEvolutionMetalpoor2019} \\   
& {[}$\alpha$/Fe{]}             & Normal       & $0.40 \pm 0.1$    & \cite{rainChemicalEvolutionMetalpoor2019}                                               \\
& $\mu$ & Uniform & $13.8 \sim 14.1$& \cite{huangMilkyWayTomography2019}\\
&&&&\cite{kaluznyClustersAgeSExperiment2014}\\
&&&&\cite{rozyczkaClustersAgeSExperiment2013}\\
& E(V-I) & Uniform & $0.08 \sim 0.15$& \cite{kaluznyClustersAgeSExperiment2014}\\
&&&&\cite{richterStromgrenPhotometryGlobular1999}\\
&&&&\cite{lallementUpdatedGaia2MASS3D2022}\\
\hline
NGC 7078 (M15) & {[}Fe/H{]}                    & Normal       & $-2.27 \pm 0.10$     &  \cite{kirbyMetallicityAlphaElementAbundance2008}\\
& & & & \cite{vandenbergConstraintsDistanceModuli2016} \\ 
& & & & \cite{leeSEGUEStellarParameter2011} \\
& {[}$\alpha$/Fe{]}             & Normal       & $0.30 \pm 0.1$    & \cite{leeSEGUEStellarParameter2011}                                               \\
& & & &  \cite{kirbyMetallicityAlphaElementAbundance2008}\\
& & & &  \cite{feuilletBVIPhotometryRed2014}\\
& $\mu$ & Uniform & $15.3 \sim 15.6$ & \cite{bhardwajOpticalInfraredPulsation2021}\\
& & & &  \cite{feuilletBVIPhotometryRed2014}\\
& E(V-I) & Uniform & $0.08 \sim 0.15$ & \cite{dutraForegroundBackgroundDust2000}\\
\hline
NGC 7099 (M30) & {[}Fe/H{]}                    & Normal       & $-2.31 \pm 0.10$     &  \cite{kainsEstimatingParametersGlobular2013}\\
& & & & \cite{sandquistWideFieldCCDPhotometry1999} \\ 
& {[}$\alpha$/Fe{]}             & Normal       & $0.40 \pm 0.1$    & \cite{carrettaNaOAnticorrelationHB2009}                                               \\
& $\mu$ & Uniform & $14.6 \sim 14.9$ & \cite{kainsEstimatingParametersGlobular2013}\\
& & & & \cite{piottoDeepLuminosityFunction1990} \\ 
& E(V-I) & Uniform & $0.0 \sim 0.10$ & \cite{piottoDeepLuminosityFunction1990}\\
\hline
NGC 4147  & {[}Fe/H{]}                    & Normal       & $-1.82\pm 0.10$     &  \cite{villanovaSpectroscopicStudyGlobular2016}\\
& {[}$\alpha$/Fe{]}             & Normal       & $0.38 \pm 0.10$    & \cite{villanovaSpectroscopicStudyGlobular2016}                                               \\
& $\mu$ & Uniform & $16.2 \sim 16.5$& \textsuperscript{\textdagger}\\
& E(V-I) & Uniform & $0.0 \sim 0.03$& \textsuperscript{\textdagger}\\
\hline
NGC 5053  & {[}Fe/H{]}                    & Normal       & $-2.25\pm 0.12$     &  \cite{sbordoneChemicalAbundancesGiant2015}\\
& & & & \cite{bobergChemicalAbundancesNGC2015} \\ 
& & & & \cite{tangMetalpoorNonSagittariusGlobular2018} \\ 
& & & & \cite{chunExtratidalStarsChemical2020} \\ 
& {[}$\alpha$/Fe{]}             & Normal       & $0.35 \pm 0.11$    & \cite{chunExtratidalStarsChemical2020}                                               \\
& & & & \cite{bobergChemicalAbundancesNGC2015} \\ 
& & & & \cite{tangMetalpoorNonSagittariusGlobular2018} \\ 
& $\mu$ & Uniform & $16.1 \sim 16.4$& \textsuperscript{\textdagger}\\
& E(V-I) & Uniform & $0.0 \sim 0.03$& \textsuperscript{\textdagger}\\
\hline
NGC 5466  & {[}Fe/H{]}                    & Normal       & $-2.00\pm 0.15$     &  \cite{lambChemicalAbundancesGlobular2015}\\
& {[}$\alpha$/Fe{]}             & Normal       & $0.21 \pm 0.11$    & \cite{lambChemicalAbundancesGlobular2015}\\
& $\mu$ & Uniform & $15.95 \sim 16.25$& \textsuperscript{\textdagger}\\
& E(V-I) & Uniform & $0.0 \sim 0.03$& \textsuperscript{\textdagger}\\
\hline
NGC 104 (47 Tuc)  & {[}Fe/H{]}                    & Normal       & $-0.77 \pm 0.05$     &  \cite{corderoDetailedAbundancesLarge2014}\\
& & & & \cite{kochNewAbundanceScale2008} \\ 
& & & & \cite{wangSodiumAbundancesAGB2017} \\ 
& {[}$\alpha$/Fe{]}             & Normal       & $0.31 \pm 0.05 $    &  \cite{corderoDetailedAbundancesLarge2014}\\
& & & & \cite{kochNewAbundanceScale2008} \\ 
& $\mu$ & Uniform & $13.15 \sim 13.45$& \cite{denissenkovConstraintsDistanceModuli2017}\\
& & & & \cite{brogaardAge47Tuc2017} \\ 
& E(V-I) & Uniform & $0.0 \sim 0.05$& \cite{denissenkovConstraintsDistanceModuli2017}\\
\hline
NGC 6362  & {[}Fe/H{]}                    & Normal       & $-1.08 \pm 0.05$     &  \cite{massariChemicalCompositionLowmass2017}\\
& & & & \cite{mucciarelliNGC6362Least2016} \\ 
& {[}$\alpha$/Fe{]}             & Normal       & $0.32 \pm 0.03 $    &  \cite{massariChemicalCompositionLowmass2017}\\
& $\mu$ & Uniform & $14.40 \sim 14.80$& \cite{olechClustersAgeSExperiment2001}\\
& E(V-I) & Uniform & $0.05 \sim 0.10$& \cite{kovacsEmpiricalRelationsCluster2001}\\
\hline
\end{tabular}
}

\vspace{1ex}
{\footnotesize \textit{Note}\textsuperscript{\textdagger}: \cite{harrisCatalogParametersGlobular1996}, \cite{carrettaIntrinsicIronSpread2009}, \cite{schlaflyMEASURINGREDDENINGSLOAN2011}, \cite{vandenbergConstraintsDistanceModuli2018} and \cite{baumgardtAccurateDistancesGalactic2021} are referenced for all GCs studied in this paper}
\vspace{0.3cm}
\end{table*}

We use the Dartmouth Stellar Evolution Program (DSEP) \citep{dotterDartmouthStellarEvolution2008} to generate stellar models and generally use literature estimates when adopting uncertainties for each parameter (see Table~\ref{tab_uni_MC} and Table~\ref{tab_indi_MC}, and discussion in \citet{yingAbsoluteAgeNGC2024}). We generate $10,000$ sets of input parameters for each GC by performing Monte Carlo simulations based on their associated probability distribution functions. Each set of input parameters is used to evolve $13$ low-mass stellar models with mass from $0.2\,M_{\odot}$ to $0.68\,M_{\odot}$ with an increment of $0.04 \,M_{\odot}$, $14$ medium-low-mass stellar models with mass from $0.7\, M_{\odot}$ to $1.35 \,M_{\odot}$ with an increment of $0.05 \,M_{\odot}$, $6$ medium-high-mass stellar models with mass from $1.4\, M_{\odot}$ to $1.9 \,M_{\odot}$ with an increment of $0.1 \,M_{\odot}$, and $6$ high-mass stellar models with mass from $2.0\, M_{\odot}$ to $3.0 \,M_{\odot}$ with an increment of $0.2 \,M_{\odot}$. The lower-mass models use the FreeEOS-2.2.1 \citep{irwinFreeEOSEquationState2012}, while the higher-mass models use an analytical equation of state which includes the Debye-Huckel correction \citep{chaboyerOPALEquationState1995}. \citet{dotterMESAISOCHRONESStelLAR2016} describes a robust method to transform a set of stellar evolution tracks onto a uniform basis and then interpolate within that basis to construct stellar isochrones. We adopt this equivalent evolutionary phase(EEP)--based method to generate $41$ theoretical isochrones from $8\,$Gyr\footnote{For more metal-rich GCs such as 47 Tuc and NGC 6362, we also construct isochrones from $6$ Gyr to $8$ Gyr.} to $16\,$Gyr with an increment of $200\,$Myr. 
Each isochrone is constructed with a dense grid of $400$ EEPs to ensure that the output isochrones have a high density of points to avoid any interpolation errors when constructing simulated color-magnitude diagrams (sCMDs).

\section{Isochrone fitting} \label{iso_fitting}

A stellar evolution code produces output at fixed points in time—timesteps—that allow one to follow the evolution of a model star over some portion of its lifetime \citep{dotterMESAISOCHRONESStelLAR2016}. The resulting model contains information such as the radius, mass, and luminosity of a star corresponding to its different evolutionary stages. However, it is very difficult to compare stellar evolution models with an observed star directly, as the number of stellar evolution parameters exceeds the observed parameters, leading to overfitting. GCs, conversely, can be modeled as simple stellar populations, as stars in a GC are believed to have originated from the same source and have similar compositions and ages. Therefore, stars in a GC provide sufficient statistical significance to be used to fit isochrones. 

We can convert theoretical surface temperature and luminosity predictions into observable quantities, such as color and magnitude, by applying bolometric correction tables. After correcting for interstellar reddening and using the distance modulus of the cluster, we can plot theoretical isochrones with different input parameters and ages on the same color-magnitude diagram (CMD) as observational data.

Several methods have been developed for fitting theoretical isochrones to CMDs. Most methods aim to detect morphological changes in the isochrones and compare them to observational data. Visual examination remains widely used; for instance, \citet{dotterACSSurveyGalactic2010} estimates the age of a given GC by determining the isochrone that best fits the CMD from the main sequence turn-off (MSTO) through the subgiant branch (SGB). This method can only be applied with a strong assumption on the underlying physics in the model (hence it is not suitable for determining the absolute age) and the distance and reddening of GCs. The lack of quantification makes it very difficult to estimate the uncertainty of the age.

Other methods aim to quantify morphological differences. \citet{ibenAsymptoticGiantBranch1983}, for example, define the quantity $\delta V$, which is the difference in magnitude between different evolution stages, such as the MSTO and the horizontal branch (HB). \citet{vandenbergAges55Globular2013} refine this method by introducing a theoretical zero-age horizontal branch (ZAHB). \citet{marin-franchACSSurveyGalactic2009} uses a similar approach with accurate determination of MSTO magnitude. Alternative methods quantify this morphological difference by defining the quantity $\delta (B-V)$, which is the difference in color between different evolution stages such as MSTO and the base of the giant branch (GB) \citep{sarajediniNewAgeDiagnostic1990}. \citet{chaboyerAccurateRelativeAge1996} improve the performance of fitting by fitting the SGB instead. Those methods extend from the visual examination method as there is an exact definition of different evolution stages in stellar evolution models, which are good theoretical calibrations. However, those stellar evolution stages are not as straightforward as those in the CMD for observational data and theoretical models. More importantly, it is very hard to quantify the performance of those metrics when they only utilize a very limited part of both observational data and . Those kinds of metrics intrinsically give up the information provided by the stars, which are not exactly at particular stages of evolution.

\begin{figure}[!htbp]
    \centering
    \includegraphics[width=0.45\textwidth]{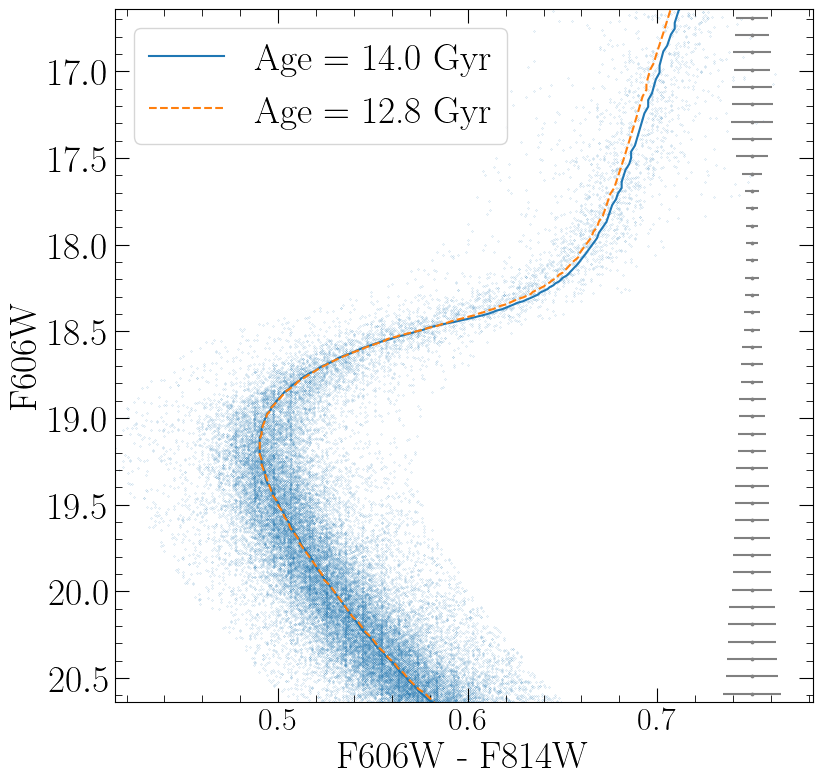}
    \caption{Comparing two theoretical isochrones we constructed with different ages, MC parameters, distance, and reddening. The observational data for M15 is plotted in the background, with photometric uncertainty estimated from the AS test plotted on the right.}
    \label{fig:M15_degen}
\end{figure}

Luminosity functions (LF) can also be used to determine the age of the GC. The distribution of the number of stars in luminosity bins will change as more massive stars evolve more rapidly than less massive ones. \citet{jimenezNewSelfconsistencyCheck1996} uses GB LF and estimated HB masses to accurately determine the age. \citet{paustBVIPhotometryLuminosity2007} also uses the LF method to determine the age of M92. Because the LF method is based on counting stars only in magnitude bins, it is relatively insensitive to the mixing-length and reddening \citep{jimenezGlobularClusterAges1998}. 

The challenge becomes more pronounced when attempting to make accurate absolute age measurements using Monte Carlo (MC) methods. Fig.~\ref{fig:M15_degen} illustrates the problem, as the two isochrones with different ages, input MC parameters, distances, and reddenings are visually almost identical on the CMD. In other words, isochrones in the form of curves on the CMD are merely a low-dimensional projection (in terms of F606W and F814W magnitudes) of a high-dimensional model (including distance, reddening, mass, luminosity, temperature, etc.) and could be degenerate, as they only contain partial information from the theoretical model. Therefore, \citet{yingAbsoluteAgeM922023,yingAbsoluteAgeNGC2024} proposes improved isochrones-fitting methods that effectively transform isochrones into a distribution that utilizes more information from stellar evolution models and observation for a direct comparison with actual observational data. 

The object can be summarized as follows:

\begin{enumerate}
    \item Physics motivated: Fitting the entire CMD (color and magnitude information of stars) simultaneously with distance and reddening.
    \item Data-driven: Utilize the state-of-the-art observational data, including the photometric uncertainty of all stars and completeness from artificial star tests. 
\end{enumerate}

\begin{table}[!htbp]
\caption{sCMD construction parameters \label{tab:sCMD}}
\begin{tabular}{lll}
\hline
Cluster  & PDMF $(\alpha)$  & Binary Fractions   \\
\hline
NGC  7078 (M15) & $-1.00$ & $0.013$  \\
\hline

\hline
NGC  7099 (M30) & $-0.80$ & $0.012$  \\
\hline

\hline
NGC  6809 (M55) & $-0.83$ & $0.04$  \\
\hline

\hline
NGC  4147 & $0.03$ & $0.029$  \\
\hline

\hline
NGC  5053 & $-1.26$ & $0.073$  \\
\hline

\hline
NGC  5466 & $-1.14$ & $0.066$  \\
\hline

\hline
NGC  104 (47 Tuc)& $-0.84$ & $0.009$  \\
\hline

\hline
NGC  6362 & $-0.58$ & $0.046$  \\
\hline

\hline
\end{tabular}

\vspace{1ex}
{\footnotesize \textit{Note:} We use binary fraction from \citet{miloneACSSurveyGalactic2012}. We use PDMF from \citet{ebrahimiNewInsightStellar2020} for NGC  104, 5053, 5466, 6362, 6809, and 7099. We use PDMF from \citet{sollimaGlobalMassFunctions2017} for NGC  4147. We use PDMF from \citet{paustACSSurveyGalactic2010} for NGC  7078.}
\end{table}

To achieve this, we aim to couple theoretical isochrones with observational data by transforming isochrones into simulated color-magnitude diagrams (sCMDs), enabling a direct comparison with observed CMDs. An sCMD is constructed by randomly sampling $4,000,000$ points for each isochrone in the following steps:
\begin{enumerate}
    \item A random distance from the cluster's center is selected from the observed distribution \citep{sarajediniACSSurveyGalactic2007}.
    \item A random mass is selected using the present-day mass function (PDMF) (see Table~\ref{tab:sCMD}). This simulated star's magnitudes (F606W and F814W) are then determined from the isochrone. 
    \item The simulated star is randomly assigned to be a member of a visual binary system using the observed binary mass fraction (see Table~\ref{tab:sCMD}). 
    \item If a star is a member of a binary system, then a secondary star is created, assuming a flat secondary mass distribution with mass ratio  $q=0.5$ to $1.0$. The magnitude of the secondary star, determined from the isochrone, is added to the primary star's magnitude to represent the binary system as a single star in the photometry.
    \item Apply artificial star test result for the corresponding GC \citep{andersonACSSURVEYGLOBULAR2008} to correct for photometric uncertainty and completeness.
\end{enumerate}

The $4,000,000$ sample points represent the distribution of stars given the set of MC parameters and age, and are used to fit the observed CMD. In previous studies \cite{yingAbsoluteAgeM922023, yingAbsoluteAgeNGC2024}, several fitting methods were proposed to utilize more information from the theoretical models in CMD-fitting, and we will discuss those methods in the following few sections.

\subsection{Voronoi Binning Method}
Voronoi Binning Method was first proposed by \citet{yingAbsoluteAgeM922023} as a new isochrone-fitting method that relies on number density estimation on the CMD. Here, we briefly summarize the method:
\begin{enumerate}
    \item Partition the 2D sCMD into sub-regions using the adaptive Voronoi binning method of \citet{cappellariAdaptiveSpatialBinning2003}. The algorithm utilizes centroidal Voronoi tessellations\citep{duCentroidalVoronoiTessellations1999} to divide a region into subregions and groups the nearest points to reach a uniform number of stars in every bin while keeping the bin shape compact. The number of simulated stars assigned to each bin $E_i$ represents the number of stars expected in that bin if the underlying population of the observed GC is the same as the theoretical isochrone.
    \item Distance and reddening correction will be applied to the observed CMD to align with the sCMD
    \item Each observed star will be assigned to predetermined Voronoi bins based on its location at the corrected CMD. The observed star counts $O_i$, the number of stars in each bin.
    \item The goodness-of-fit will be determined using a $\chi^2$ metric:
    \begin{equation}
    \chi^2 = \sum_{i} \frac{\left( O_i - E_i \right)^2}{E_i} \label{eq2},
    \end{equation}
    \item A parametric bootstrapping method will be used to resample the observed data using the photometric error and completeness from the artificial star test \citep{andersonACSSURVEYGLOBULAR2008}. The resampled data will be compared with observed data using the same method to determine the intrinsic $\chi^2$ results from the photometric error and missing stars. 
    \item Empirical $\chi^2$ distribution from $10,000$ rounds of bootstrapping will be used to assign the weight of each isochrone.
    \item The final age estimation will be based on the CMD fitting and calibration star weights.
\end{enumerate}
\citet{yingAbsoluteAgeM922023} estimate the absolute age of M92 $=13.80 \pm 0.75$ Gyr using this method, which is twice as accurate as the age estimated in \citet{omalleyAbsoluteAgesDistances2017} with a similar Monte-Carlo isochrone constructing method but with only MS fitting. This demonstrates the capability of a full-CMD-fitting method. For example, the isochrones with age $=12.8$ Gyr in Fig.~\ref{fig:M15_degen} has $\chi^2/\textup{DOF} = 0.145$ while the isochrones with age $=14.0$ Gyr has $\chi^2/\textup{DOF} = 0.154$. The empirical $\chi^2$ distribution combined with the calibration star test suggests that the isochrones with age $=12.8$ Gyr are $>10$ times more likely to be a good fit compared to the isochrones with age $=14.0$ Gyr.

Because the Voronoi binning method is a full-CMD-fitting method, it is susceptible to morphological changes anywhere on the CMD. As a result, it is extremely selective. \citet{yingAbsoluteAgeM922023} shows that only $1,100$ isochrones out of the total $820,000$ isochrones generated fit the observational data. The success of this method relies heavily on the accuracy of the estimated photometric uncertainty. Even though the Voronoi bins are generated by combining the nearest neighbors, after the bins are generated, the $\chi^2$ metric from Equation~\ref{eq2} assumes independence between bins. The empirical $\chi^2$ distribution captures the intrinsic $\chi^2$ caused by such a shift, which is a relatively small contribution compared to that from the difference in the isochrone morphology as \citet{yingAbsoluteAgeM922023} generates sCMD for M92, which is almost indistinguishable from the CMD from observational data. The success of this method relies on the assumption that we have an accurate estimation of the photometric uncertainty from the artificial star test. However, this is not the case for every GC.

\begin{figure}[!htbp]
    \centering
    \includegraphics[width=0.45\textwidth]{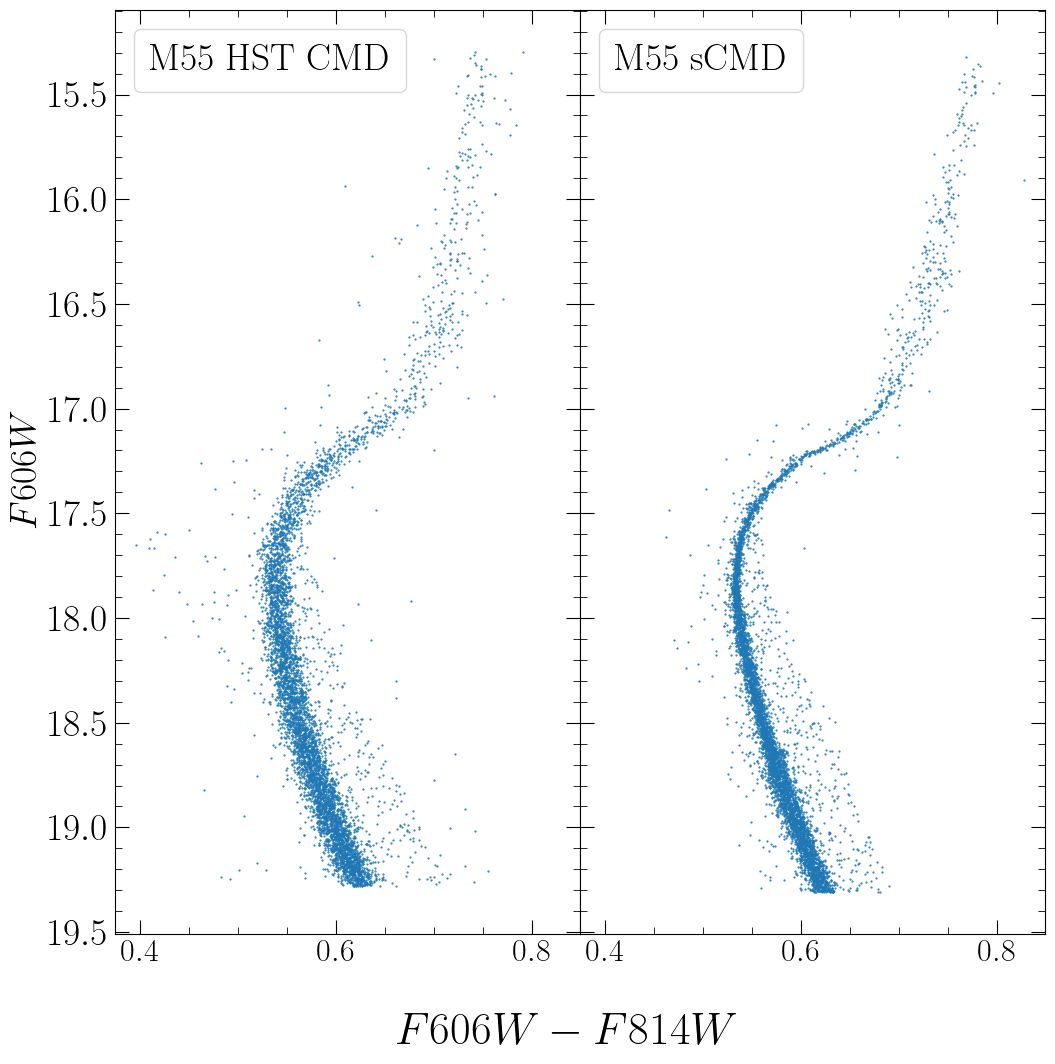}
    \caption{Comparing the CMD generated using HST ACS data for M55 (left) with the sCMD generated using photometric uncertainty estimated from the AS test for M55 (right).}
    \label{fig:M55_compare}
\end{figure}

Fig.~\ref{fig:M55_compare} compares the CMD of M55 and sCMD generated using photometric uncertainty estimated from the AS test for M55. There is a noticeable ``width" difference between those two plots as the observational data tend to have a more considerable variance in color compared to the simulated data. This is not observed in \citet{yingAbsoluteAgeM922023} with M92 data. The most likely cause for this is that the photometric uncertainty used to generate the sCMD is systematically underestimated. The ``width" difference is crucial to a full-CMD-fitting method such as the Voronoi binning method, as the intrinsic $\chi^2$ caused by photometric uncertainty may dominate the $\chi^2$ metric, and parametric bootstrapping methods will be unable to quantify it with inaccurate photometric uncertainty.

We test several other hypotheses that may explain the ``width" difference. For example, photometric binaries will be brighter compared to a single star, they appear to be further from the isochrones which leads to a wider CMD. However, increasing the binary fraction cannot resolve the problem, especially for giant branch stars, which are hardly affected. Differential reddening might be the potential reason for the wider CMD. However, we have already considered the influence of differential reddening when we select our target cluster, as we only select clusters with low reddening. In a recent study, \citet{legnardiDifferentialReddeningDirection2023} examined the effect of differential reddening on GCs. They found there is no significant differential reddening for M55. Multiple stellar populations can also cause the color ``width" to increase \citep{miloneMultiplePopulationsStar2022}. However, \citet{vandenbergModelsMetalpoorStars2022} shows that the multiple stellar populations problem will have a minimal effect in the F606W and F814W filters, which are the two main filters used from HST ACS data.
\begin{figure}[!htbp]
    \centering
    \includegraphics[width=0.45\textwidth]{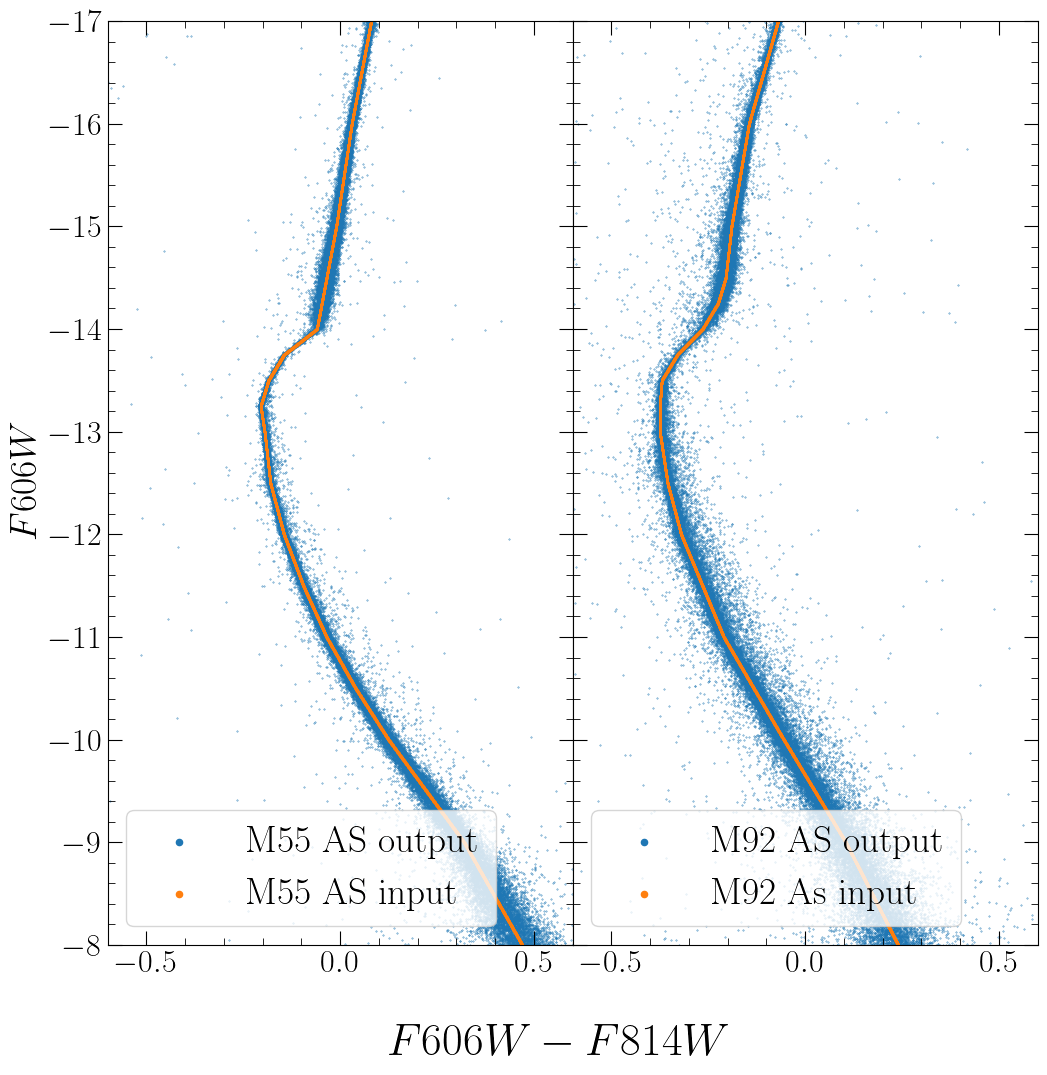}
    \caption{Comparing artificial star test results for M55 (left) with artificial star test results for M92 (right). Note that the magnitude used in the AS test is not directly comparable to observation, but it is converted when we construct sCMD.}
    \label{fig:AS_compare}
\end{figure}

\citet{andersonACSSURVEYGLOBULAR2008} state that there are two main sources of photometric error: the presence of other stars and errors in the point spread function (PSF). The algorithm used to perform the photometry can be compromised by the presence of neighbors. Fig.~\ref{fig:AS_compare} compares the results of the AS test for M92 and M55. The photometric uncertainty is estimated by measuring the difference in magnitude between input and output artificial stars. M92 has a noticeably larger photometric uncertainty because M92 is a more crowded GC than M55. Because crowding is the dominant source of uncertainty in M92, \citet{yingAbsoluteAgeM922023} can reproduce accurate sCMDs, which can be analyzed using the Voronoi binning methods. For a less crowded GC, such as M55, PSF modeling can be the dominant source of uncertainty. In fitting artificial stars, the same PSF was used as that used to create them, while the PSF model can differ at different exposures due to spacecraft breathing or focus changes. Therefore, we cannot expect the same perfect match of the PSF with the real PSF of real stars \citep{miloneACSSurveyGalactic2012}. Unfortunately, it is impossible to quantify the uncertainty due to the PSF modeling without redesigning the AS test process.

\subsection{2D Kolmogorov–Smirnov method}
\citet{yingAbsoluteAgeNGC2024} develop a testing method that is more robust to the uncertainty while maintaining the advantage of a full-CMD-fitting method. 2D Kolmogorov–Smirnov (KS) test is the multivariate extension of the famous Kolmogorov–Smirnov test and a goodness-of-fit test that can be used to test for consistency between the empirical distribution of data points on a plane and a hypothetical density law \cite[e.g.][]{peacockTwodimensionalGoodnessfitTesting1983, fasanoMultidimensionalVersionKolmogorovSmirnov1987}.


The process can be summarized as follows:
\begin{enumerate}
    \item Choose a set of distance modulus and reddening values and apply correction to the observed CMD to align with the sCMD.
    \item Estimate the empirical cumulative distribution function(ECDF) of the sCMD as the expected ECDF for the isochrone using a divide and conquer algorithm \citep{bentleyMultidimensionalDivideconquer1980}. Generate a linear interpolation of the expected ECDF to cover the CMD plane.
    \item Apply the same method to determine the observed ECDF from the observed CMD.
    \item Compare observed ECDF and expected ECDF at the location of each observed star and return the maximum difference.
    \item Predict the next set of distance modulus and reddening values based on the Gaussian process within the predetermined boundary values, and rerun the analysis process.
\end{enumerate}

Compared to the Voronoi binning method, the 2D KS method is more robust to the underestimated photometric uncertainty because it uses the cumulative distribution of stars on the CMD plane rather than assuming independence between sub-regions on the CMD plane. As a result, the ``width problem" introduced by unknown uncertainties, as shown in Fig.~\ref{fig:M55_compare}, is attenuated by this method's ``cumulative" nature. Furthermore, as a non-parametric method, we can avoid the process of hyperparameter tuning, which is necessary for the Voronoi binning method, where we need to set the optimal bin numbers and signal-to-noise ratio to balance accuracy and computational cost. 

Similar to the Voronoi binning method, the 2D KS method does not have a closed-form distribution of samples assuming the null hypothesis. More importantly, \citet{babuAstrostatisticsGoodnessFitAll2006} suggests that the distribution of KS statistics varies with the underlying true distribution for a multivariate KS test. As a result, the 2D KS statistics studies by several astronomical literature \cite[e.g.][]{peacockTwodimensionalGoodnessfitTesting1983, fasanoMultidimensionalVersionKolmogorovSmirnov1987} cannot be applied directly to our case. To make the 2D KS method a test statistic, \citet{yingAbsoluteAgeNGC2024} suggests a method based on bootstrap resampling to determine the empirical distribution of the 2D KS metric under the null hypothesis. Here, we briefly summarize the process:
\begin{enumerate}
    \item Set an isochrone with the lowest 2D KS test metric as the base model.
    \item Generate a set of simulated observational data using the base model on the sCMD plane. 
    \item Construct $10,000$ resamples based on the same parametric model for the target GC
    \item Compare simulated observed ECDF and simulated expected ECDF at the location of each observed star and return the maximum difference.
    \item The final distribution of the 2D KS test statistics will be used as testing statistics.
\end{enumerate}

We select two methods based on the number of members in each GC. We only use the Voronoi Binning method for M15, M30, and 47 Tuc as they have a large number of members and more significant photometric uncertainties due to crowding. We use the 2D KS test method for M55, NGC 4147, NGC 5053, NGC 5466, and NGC 6362 as they have fewer members and are more sensitive to the unidentifiable photometric uncertainty. We also use the 2D KS test method for comparison for M15, M30, and 47 Tuc.

\subsection{Multiple populations in GCs}

\begin{table*}[!htbp]
\caption{Best-fit Parameters Derived from Fitting Isochrones to the Fiducial Lines Derived from the 47 Tuc Photometry \label{tab:2pop}}
\begin{tabular}{llllllll}
\hline
Population  & Age   & Distance Modulus & Reddening & [Fe/H] & He Abundance & [$\alpha$/Fe]  \\
\hline
P1 & $11.91 \pm 0.86$ & $13.39 \pm 0.04$  & $0.01 \pm 0.01$ & $-0.78 \pm 0.04$ & $0.253 \pm 0.002$ & $0.20 \pm 0.02$\\
\hline

\hline
P2 & $12.02 \pm 0.87$ & $13.36 \pm 0.05$  & $0.02 \pm 0.01$ & $-0.79 \pm 0.04$ & $0.254 \pm 0.002$ & $0.24 \pm 0.08$\\
\hline

\hline
\end{tabular}
\end{table*}

Although GCs were historically considered to be single populations of coeval stars and have served as the ideal laboratory for stellar evolution model calibration, recent studies have shown most GCs host multiple populations with variance in both light-element abundance \cite[e.g.][]{ carrettaNaOAnticorrelationHB2009,miloneHubbleSpaceTelescope2018,alvarezgarayMgKAnticorrelationCentauri2022} and metallicity \cite[e.g.][]{yongConfirmingIntrinsicAbundance2016, lardoConfirmationMetallicitySpread2022, marinoMetallicityVariationsChromosome2023} which results in the splitting/broadening along MS, SGB, and extent to RGB \cite[e.g.][]{cassisiDoubleSubgiantBranch2007,piottoTripleMainSequence2007}.

When modeling data with multiple subpopulations using a single-population model, the model may misestimate the uncertainty of its predictions as the single-population model tends to overestimate the variance of the data to fit a single broad distribution to multiple different subpopulations. Therefore, it is essential to examine the potential impact of multiple populations within the current framework. It is worth noting that most studies on multiple populations in GCs were conducted using filters much bluer (e.g., UV filters in HST) than the F606W and F814W used in this paper, where \citet{vandenbergModelsMetalpoorStars2022} suggests that this multiple population is hardly observable in red filters.

The primary challenge in directly modeling multiple populations is the computational complexity. Absolute age determinations of GCs require extensive computational resources, as millions of stellar evolution models must be generated to adequately sample the parameter space. Since it will take $\sim 1$ hour to evolve a stellar evolution model, a fully Bayesian approach, such as Markov Chain Monte Carlo (MCMC), would be computationally prohibitive due to the extremely long chains required. Instead, we adopt a Monte Carlo approach that operates asynchronously to enhance efficiency. To model multiple populations, we want to use a hierarchical Bayesian model to fine-tune the second population based on the model for the first population. As a result, we cannot use an asynchronous approach, making it practically infeasible to implement. We are developing the next generation of stellar evolution models based on a generative machine learning architecture, which aims to significantly reduce the computational cost associated with evolving stellar models and overcome this difficulty \citep{ying2025}.

\citet{boudreauxChemicallySelfconsistentModeling2025} describes a method to study multiple populations in NGC 2808. Instead of constructing sCMDs from models and fitting the observed CMDs, we use fidanka \citep{boudreaux_fidanka_2023} to extract multiple isochrones from observed CMDs and compare them with theoretical isochrones. Although this method is based on a different approach compared to the rest of the paper, it significantly reduces computational complexity from quadratic to linear. It helps us to gain insights into the implications of multiple populations in this study.

\begin{figure}[!htbp]
    \centering
    \includegraphics[width=0.45\textwidth]{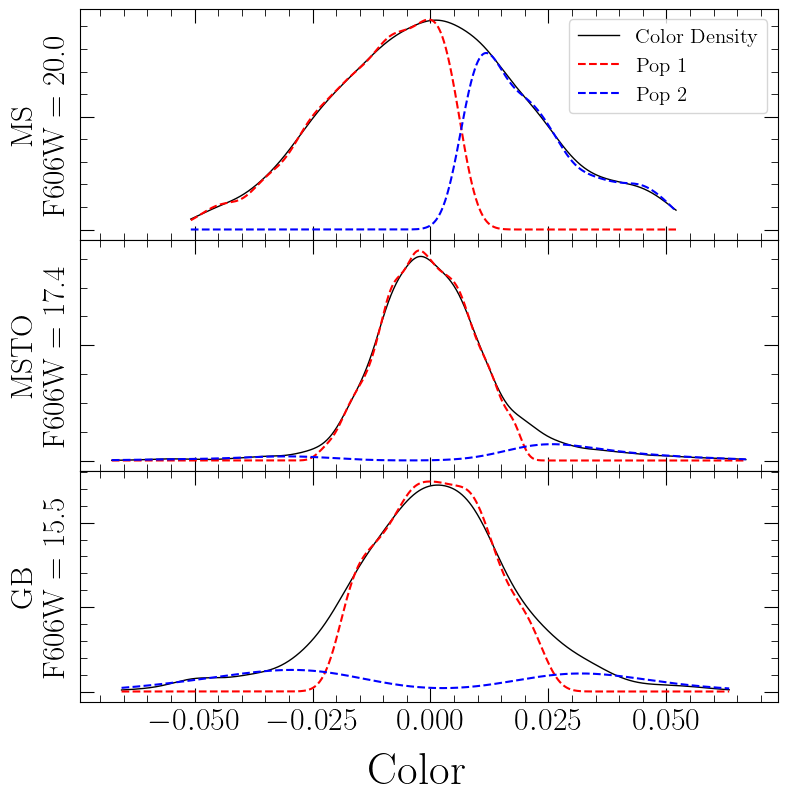}
    \caption{Examples of color-density profiles for three different evolution stages: Main Sequence (top), Main Sequence Turn Off (mid), and Giant Branch (bottom) on the CMD. The X-axis is the difference in color from the median ridgeline.}
    \label{fig:Color_Density}
\end{figure}

We use 47 Tuc as our test case, as it shows the strongest evidence for multiple populations around the turn-off region among the clusters we study in this paper. Previous studies have shown that multiple populations exist in 47 Tuc \cite[e.g.][]{andersonMixedPopulationsGlobular2009, miloneJWSTProject472025}. We implement the Dirichlet process \citep{fergusonBayesianAnalysisNonparametric1973, pedregosaScikitlearnMachineLearning2011}, a non-parametric Bayesian method, in the Bayesian Gaussian Mixture Model (BGMM) to determine the optimal number of subpopulations. This method enables dynamic variation in the number of inferred populations from one magnitude bin to another. Fig.~\ref{fig:Color_Density} shows the examples of color-density profiles for three different evolution stages. It suggests that the main population contributes more than $80\% $ of the total population in more than $80\%$ of the magnitude bins. The second population is only significant in the lower main sequence region, as shown in the top panel in Fig.~\ref{fig:Color_Density}. However, two strongly skewed Gaussians, positioned on opposite sides of the preference scale, will combine to form a distribution nearly indistinguishable from a single Gaussian. Thus, observing such a shape is not, by itself, strong evidence for two underlying populations. Binary stars appear brighter than single stars due to the combined luminosity of both components, naturally broadening the main sequence observed in the CMD. Additionally, the morphology of main-sequence stars of GCs on the CMD suggests that binary systems tend to shift toward bluer colors, thereby contributing to the blue tail feature illustrated in the top panel of Fig.~\ref{fig:Color_Density}. The age-sensitive features are in the MSTO and sub-giant branch, so it is unlikely that a multiple population would impact our age determination.

\begin{figure}[!htbp]
    \centering
    \includegraphics[width=0.45\textwidth]{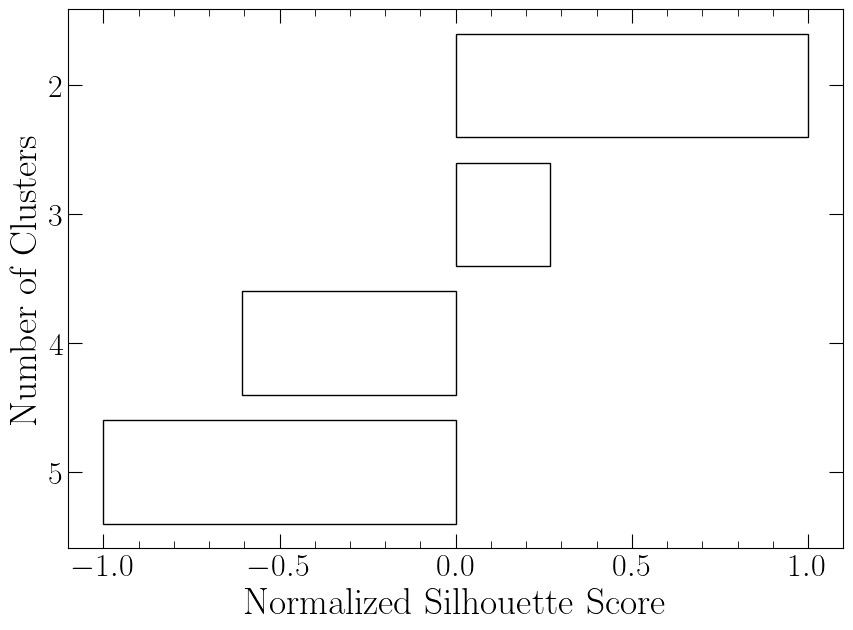}
    \caption{Sihouette analysis for 47 Tuc. The Silhouette scores represent the average score within each magnitude bin. Scores have been normalized, with values ranging from $+1$ (well-separated hypotheses) to $-1$ (poorly separated hypotheses).}
    \label{fig:silhouette}
\end{figure}

We also use the silhouette analysis \citep{rousseeuwSilhouettesGraphicalAid1987,shahapureClusterQualityAnalysis2020} method described in \citet{boudreauxChemicallySelfconsistentModeling2025} to examine the possibility for more than two populations. Silhouette analysis can be used to evaluate the quality of clustering. It provides a measure of how similar an object is to its own cluster compared to other clusters, naturally penalizing over-clustering. Since silhouette analysis requires at least 2 clusters, we determine the average silhouette score for the hypotheses that there are two, three, four, or five populations in each magnitude bin as shown in Fig.~\ref{fig:silhouette}. Silhouette analysis shows a stronger preference for two subpopulations than for more subpopulations. Therefore, we assume at most two subpopulations coexist in 47 Tuc.

Fidanka takes an iterative approach to extract fiducial isochrones from observed CMD. Fidanka divides the CMD into $100$ magnitude bins based on the F606W magnitude and uses the inverse of the convex hull area formed by $50$ nearest neighbors on the CMD to estimate the density of each star on the CMD. Fidanka assumes a Gaussian mixing model with photometric uncertainty as the variance and uses the Bayesian Gaussian mixture modeling method to cluster stars in each magnitude bin into two subpopulations.

\begin{figure}[!htbp]
    \centering
    \includegraphics[width=0.45\textwidth]{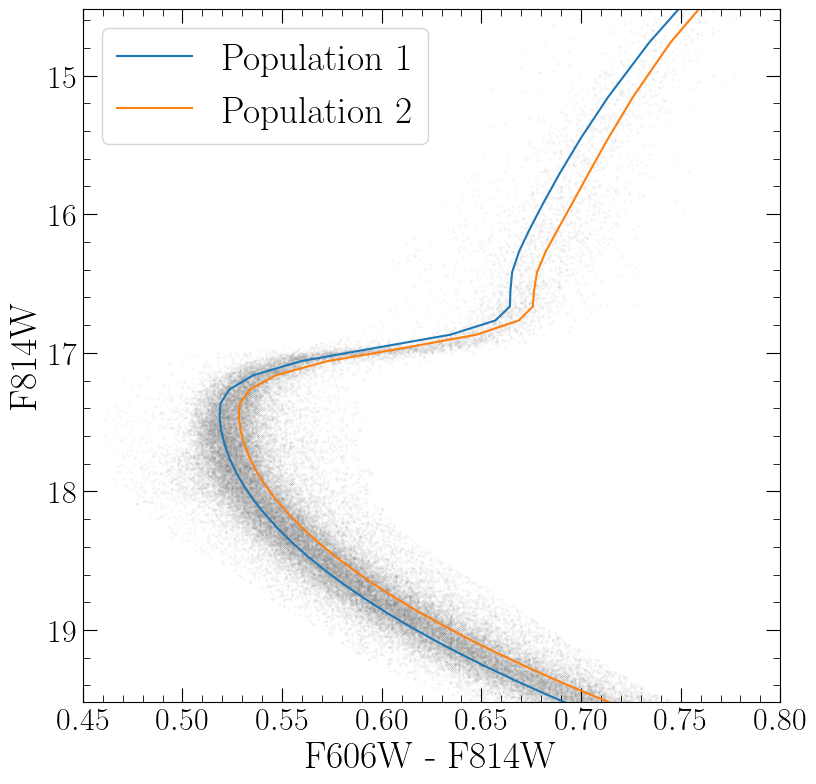}
    \caption{Fiducial isochrones calculated for two subpopulations for 47 Tuc.}
    \label{fig:47 Tuc_2pop}
\end{figure}

Fig.~\ref{fig:47 Tuc_2pop} shows the two fiducial isochrones for 47 Tuc extracted using fidanka, each potentially representing a distinct stellar subpopulation. However, we emphasize that the current filter set provides little evidence supporting the existence of two separate populations. We fit each fiducial isochrone with theoretical isochrones for 47 Tuc using a $\chi^2$ distance:
\begin{equation}
    \chi^2 = \sum \sqrt{\Delta \textup{color}^2 + \Delta \textup{mag}^2},
\end{equation}

where $\Delta \textup{color}$ is the difference in F606W - F814W and $\Delta \textup{mag}$ is the difference in F606W. Table~\ref{tab:2pop} presents the best-fit parameters derived from fitting isochrones to 2 fiducial lines derived from the 47 Tuc photometry, respectively. We confirm that F606W and F814W are not ideal filters for studying the multiple populations in GCs. Uncertainties from external parameters such as distance or reddening will dominate the internal differences between multiple populations of stars. Previous studies \cite[e.g.][]{boudreauxChemicallySelfconsistentModeling2025,dotterStellarModelsMultiple2015} suggest that self-consistent isochrones give very similar ages for the extreme multiple populations in GCs, which are followed by our study as the age of two subpopulations of 47 Tuc are consistent within their respective uncertainties with the absolute age of 47 Tuc without multiple population assumption (see section \ref{age_estimation}). \citet{miloneHubbleSpaceTelescope2018} suggests a small variance in the helium abundance between 2 populations of stars in 47 Tuc, which aligns with our result. The same analysis was applied to all GCs studied in this project, yielding similar outcomes. We confirm \citet{vandenbergModelsMetalpoorStars2022}'s statement that F606W and F814W are not ideal filters for studying the multiple populations in GCs.

\section{Results} \label{results}
\subsection{Age Estimation} \label{age_estimation}

\begin{table*}[!htbp]
\caption{Summary of results \label{tab:results}}
\centering
\resizebox{0.9\textwidth}{!}{
\begin{tabular}{llllll}
\hline
\hline
Cluster  & Age (Gyr)  & [Fe/H]  & Distance  Modulus  & Reddening & Method (Source) \\
\hline
\hline
NGC  7078 (M15) & $13.23 \pm 0.51$ & $-2.31 \pm 0.09$ & $15.45 \pm 0.03$ & $0.08 \pm 0.01$ & Voronoi Binning \\

 & $13.39 \pm 0.39$ & $-2.30 \pm 0.07$ & $15.49 \pm 0.03$ & $0.08 \pm 0.01$ & 2D KS \\
\hline
& $13.0 \pm 1.0$ & & & & \citep{feuilletBVIPhotometryRed2014}\\
& $12.75 \pm 0.25$ & & & & \citep{vandenbergAges55Globular2013}\\

\hline
\hline
NGC  7099 (M30) & $12.62 \pm 0.59$ & $-2.23 \pm 0.11$ & $14.82 \pm 0.05$ & $0.03 \pm 0.01$ & Voronoi Binning \\

 & $12.57 \pm 0.54$ & $-2.20 \pm 0.10$ & $14.81 \pm 0.06$ & $0.03 \pm 0.01$ & 2D KS \\
\hline
& $12.80 \pm 0.17 \pm 0.8 $ & & & & \citep{gontcharovIsochroneFittingGalactic2024}\\
& $13.0 \pm 1.0 $ & & & & \citep{kainsEstimatingParametersGlobular2013}\\
\hline
\hline
NGC  6809 (M55) & $12.22 \pm 0.56$ & $-1.77 \pm 0.09$ & $14.03 \pm 0.05$ & $0.10 \pm 0.01$ & 2D KS \\
\hline

& $12.9 \pm 0.8$ & & & & \citep{vandenbergConstraintsDistanceModuli2018}\\
& $13.0 \pm 0.1 \pm 0.8$ & & & & \citep{gontcharovIsochroneFittingGalactic2023a}\\
\hline
\hline
NGC  4147 & $12.48 \pm 0.54$ & $-1.74 \pm 0.10$ & $16.48 \pm 0.03$ & $0.01 \pm 0.01$ & 2D KS \\
\hline
& $12.25 \pm 0.25 $ & & & & \citep{vandenbergAges55Globular2013}\\

\hline
\hline
NGC  5053 & $13.14 \pm 0.54$ & $-2.31 \pm 0.11$ & $16.25 \pm 0.04$ & $0.01 \pm 0.01$ & 2D KS \\
\hline
& $12.25 \pm 0.38 $ & & & & \citep{vandenbergAges55Globular2013}\\
& $12.70 \pm 0.11 \pm 0.8$ & & & & \citep{gontcharovIsochroneFittingGalactic2024}\\

\hline
\hline
NGC  5466 & $11.82 \pm 0.73$ & $-1.93 \pm 0.13$ & $16.14 \pm 0.06$ & $0.02 \pm 0.01$ & 2D KS \\
\hline
& $12.15 \pm 0.11 \pm 0.8 $ & & & & \citep{gontcharovIsochroneFittingGalactic2024}\\
& $12.50 \pm 0.25 $ & & & & \citep{vandenbergAges55Globular2013}\\
\hline
\hline
NGC  104 (47 Tuc) & $11.89 \pm 0.84$ & $-0.78 \pm 0.04$ & $13.37 \pm 0.05$ & $0.02 \pm 0.01$ & Voronoi Binning\\

 & $11.76 \pm 0.76$ & $-0.77 \pm 0.03$ & $13.40 \pm 0.05$ & $0.02 \pm 0.01$ & 2D KS\\

 & $11.36 \pm 0.81$ & $-0.77 \pm 0.05$ & NA & NA & DEBs\\
\hline
& $10.4 \sim 13.4 $ & & & & \citep{brogaardAge47Tuc2017}\\
& $12.42 \pm 0.05 \pm 0.08$ & & & & \citep{simunovicGeMSGSAOIGalactic2023}\\
\hline
\hline
NGC  6362 & $11.61 \pm 0.98$ & $-1.09 \pm 0.04$ & $14.69 \pm 0.06$ & $0.06 \pm 0.01$ & 2D KS \\

 & $11.83 \pm 0.52$ & $-1.07 \pm 0.05$ & NA & NA & DEBs\\

\hline
& $12.4 \pm 0.8$ & & & & \citep{vandenbergConstraintsDistanceModuli2018}\\
& $12.5 \pm 0.5$ & & & & \citep{dotterACSSurveyGalactic2010}\\
& $12.0 \pm 0.1 \pm 0.8 \pm 0.5$ & & & & \citep{gontcharovIsochroneFittingGalactic2023}\\
\hline

\hline
\hline
\end{tabular}
}
\end{table*}

\begin{figure*}[!htbp]
    \centering
    \includegraphics[width=0.8\textwidth]{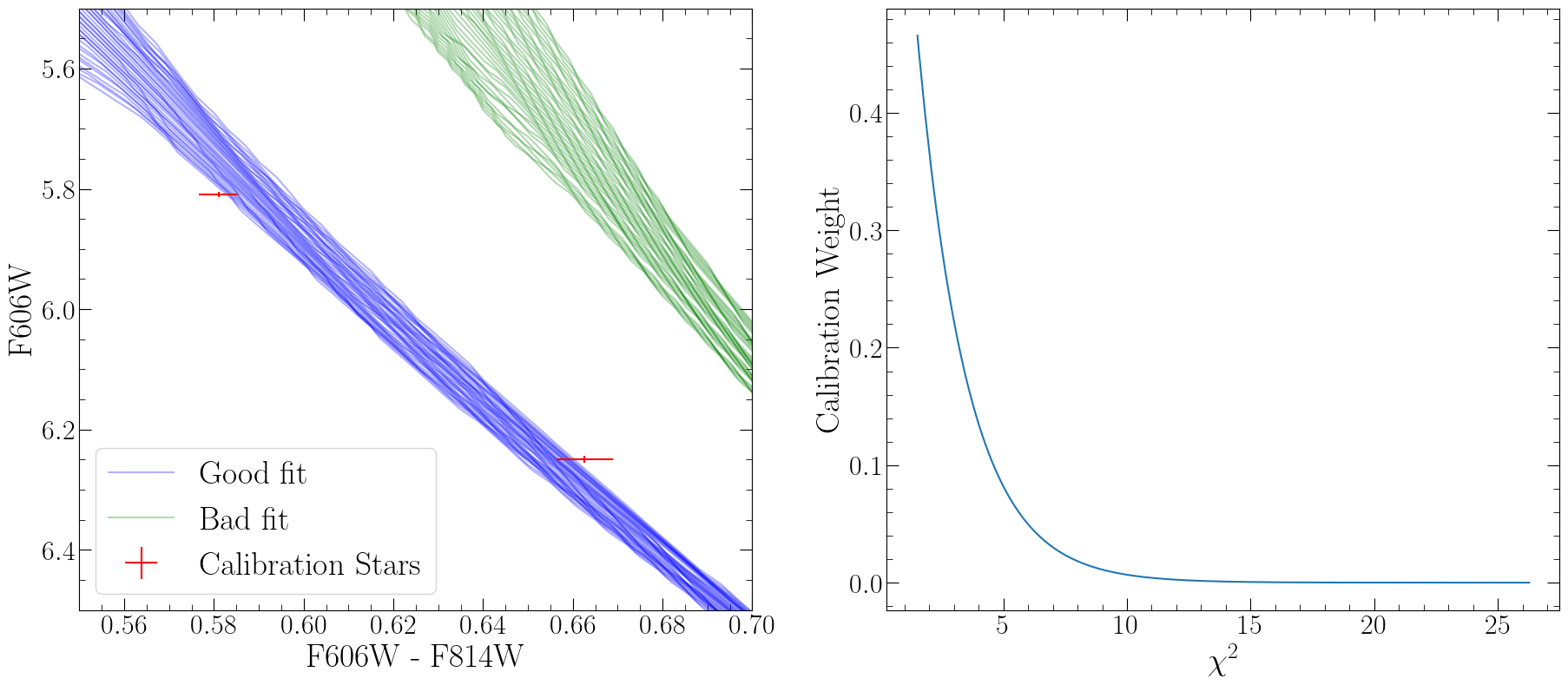}
    \caption{Left (a): An example of the Calibration star test on CMD for M55. $2$ calibration stars are shown in red with uncertainties. The MC parameters used to generate the set of blue isochrones are also considered calibrated by those two stars, but the MC parameters used to generate the set of green isochrones are not.
    Right (b): The Calibration weight is assigned according to the $\chi^2$ values.}
    \label{fig:M55_cali}
\end{figure*}

To estimate the absolute age of GCs, we combine the results from Calibration star tests and full-CMD fitting methods. Here, we use M55 as an example. For M55, we select $2$ calibration stars: HD103269  and HD108200, which have very similar compositions as M55. As a result, if the correct physics is applied, as described by the set of MC parameters for M55, we should be able to evolve star models that are very close to the calibration stars on the CMD. We perform a $\chi^{2}$ goodness-of-fit test between the two calibrating stars and each age of every MC isochrone set. The $\chi^{2}$ metric is defined as:
\begin{equation}
    \chi^2_{\textup{iso}} = \frac{\left(\textup{V}_{\textup{iso}} -  \textup{V}_{\textup{Cali}}\right)^2}{\sigma^2_{\textup{V}, \textup{Cali}}} + \frac{\left(\textup{VI}_{\textup{iso}} -  \textup{VI}_{\textup{Cali}}\right)^2}{\sigma^2_{\textup{VI}, \textup{Cali}}},
\end{equation}
where $V$ is the magnitude in the F606W band and $VI$ is the difference between the magnitude in the F606W band and the F814W band. Since the ages of the calibration stars were unknown, we only selected the isochrone with the lowest $\chi^2$ value to represent the goodness-of-fit for the set of isochrones with the same MC parameters. 

Fig.~\ref{fig:M55_cali}(a) compares two sets of isochrones with the difference mainly in $[\alpha/\textup{Fe}]$. The set of isochrones in blue ($[\alpha/\textup{Fe}] = 0.4$) are considered well-calibrated with the best fit $\chi^2 = 1.529$, which can be converted to the calibration weight $ = 0.47$ using the $\chi^2$ distribution. The set of isochrones in green ($[\alpha/\textup{Fe}] = 0.0$) is not considered well-calibrated with the best fit $\chi^2 = 26.264$, which can be converted to the calibration weight $ = 0$. Fig.~\ref{fig:M55_cali}(b) shows the relationship between $\chi^2$ and calibration weight.

For GCs where crowding is the dominant source of photometric uncertainties (i.e., M15, M30, and 47 Tuc), we expect the AS test to accurately reflect the photometric uncertainty in observations. Therefore, we apply both the 2D KS test method and the Voronoi binning method. For less populated GCs, we expect the AS test to underestimate the photometric uncertainty. Therefore, we apply the 2D KS test method solely to focus on the morphological changes in isochrones on the CMD plane.

\begin{figure}[!htbp]
    \centering
    \includegraphics[width=0.45\textwidth]{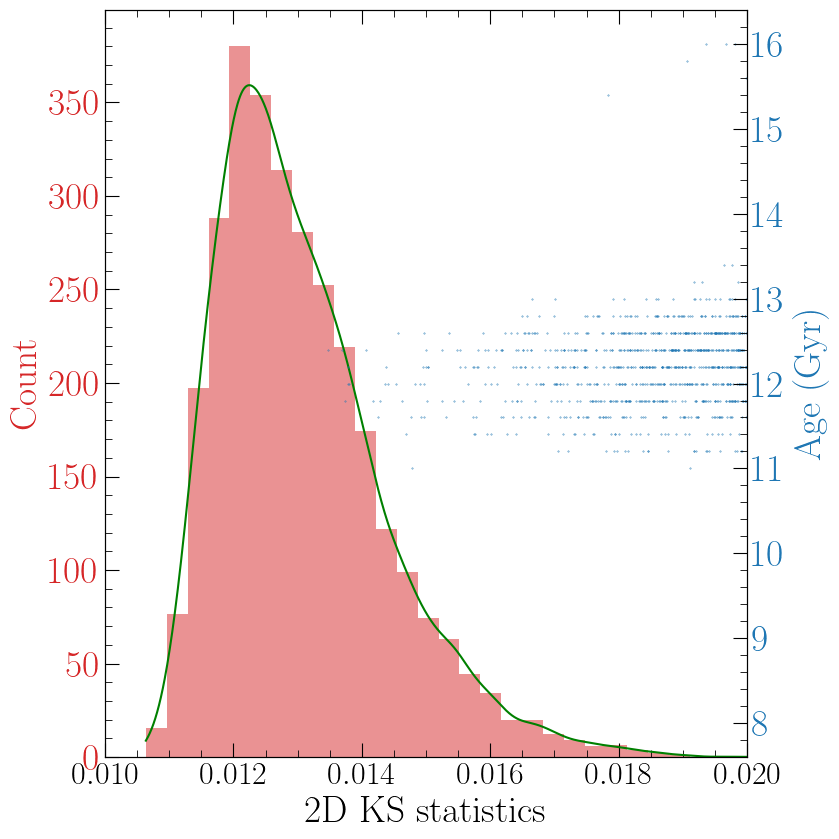}
    \caption{Bootstrap resampling for 2D-KS method. The distribution of the 2D-KS statistics from $10,000$ bootstrap resampling is shown in the grey histogram with the estimated kernel density function in red. Blue dots are the distribution of 2D-KS metric when fitting theoretical isochrones with observed data.}
    \label{fig:M55_chi2}
\end{figure}

Fig.~\ref{fig:M55_chi2} shows the 2D-KS statistics obtained through bootstrap resampling for M55 (in red), which serves as the reference probability distribution. Each of the $10,000$ sets of isochrones (in total of $420,000$ isochrones) is tested, and the distribution of results is shown in blue dots. A CMD-fitting weight: $w_{\textup{CMD}}$ is assigned to each isochrone tested, and the final weight is assigned by combining the result from the calibration star test and full CMD-fitting:
\begin{equation}
    w_{\textup{iso}} = w_{\textup{Cali}} \cdot w_{\textup{CMD}},
\end{equation}
and is used as the weight to estimate the age and other parameters. 

For GCs hosting DEBs (i.e., M55, 47 Tuc, and NGC  6809), we adopt the same isochrone fitting method described in \cite{yingAbsoluteAgeNGC2024}. We define the goodness-of-fit of $k$-th point in isochrone for $j$-th star as (where the subscripts $M$,$R$  $L$ represent mass, luminosity, and radius):
\[
\chi^2_{j,k} = \frac{\left(o_{M,j} - t_{M,k}\right)^2}{\sigma_{o_{M,j}}^2} + \frac{\left(o_{L,j} - t_{L,k}\right)^2}{\sigma_{o_{L,j}}^2} + \frac{\left(o_{R,j} - t_{R,k}\right)^2}{\sigma_{o_{R,j}}^2},
\]
For each star. For each isochrone, we find:
\begin{equation}
\chi^2_{\textup{iso}} = \sum_{\textup{star} j} \min_{k \in \textup{iso}}\left\{ \chi^2_{j,k} \right\} \label{eq:chi2_iso},
\end{equation}
which is the sum of the minimal $\chi^2$ value for each star and assigned $\chi^2_{\textup{iso}}$ as the goodness-of-fit metric for the isochrone. Bootstrap resamplings are applied to obtain the empirical distribution of $\chi^2_{\textup{iso}}$ assigned weight to isochrones based on goodness-of-fit to estimate the age and other parameters.  

\begin{figure}[!htbp]
    \centering
    \includegraphics[width=0.45\textwidth]{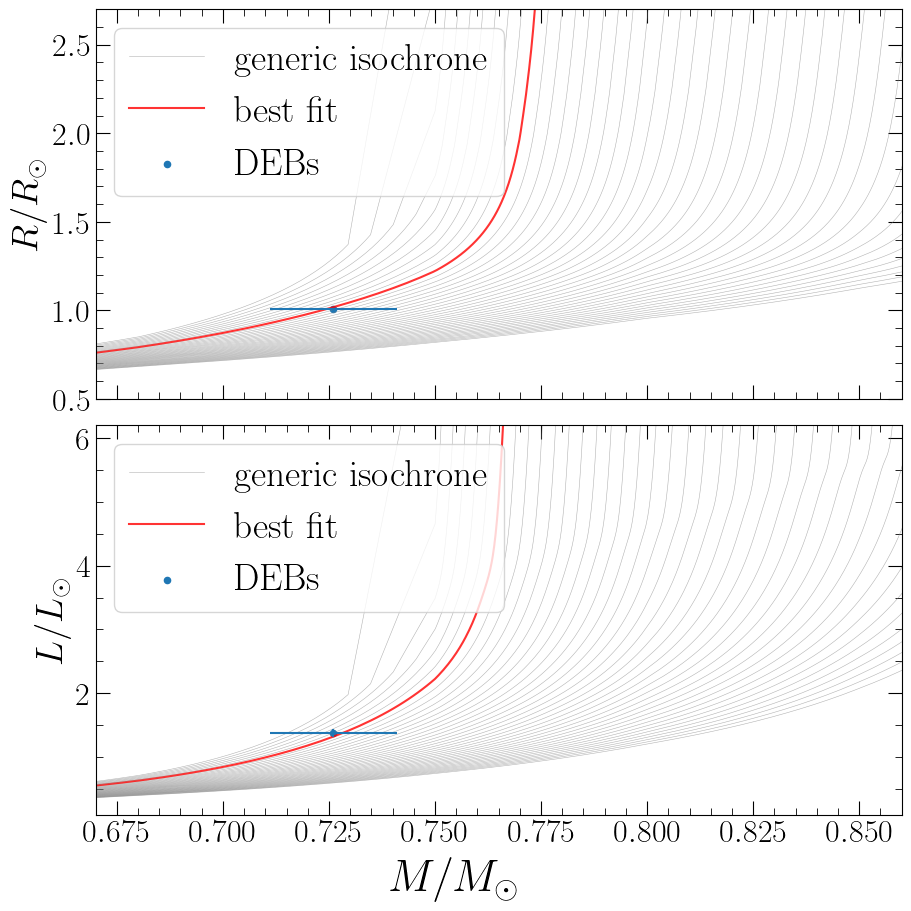}
    \caption{Comparing a set of theoretical isochrones with different ages from $8$ Gyr to $16$ Gyr on the Mass-Luminosity-Radius space to the DEB in M55. Blue points are the observational data with corresponding uncertainties. Top: isochrones and DEBs on mass vs radius plane. Bottom: Isochrones and DEBs on the mass-luminosity plane.}
    \label{fig:M55_EB}
\end{figure}

Although M55 hosts DEBs, the only set of DEBs for M55 are both main sequence stars, and their predicted properties are relatively insensitive to age. Fig.~\ref{fig:M55_EB} compares the set of DEBs for M55 with a set of theoretical isochrones on the Mass-Luminosity-Radius space with a set of theoretical isochrones with different ages from $8$ Gyr to $16$ Gyr. Because they are both main-sequence stars, with relatively considerable uncertainty in their masses, they have very low limiting effectiveness in constraining their ages. Therefore, we choose not to use the M55 DEBs as calibration stars.

\begin{figure}[!htbp]
    \centering
    \includegraphics[width=0.45\textwidth]{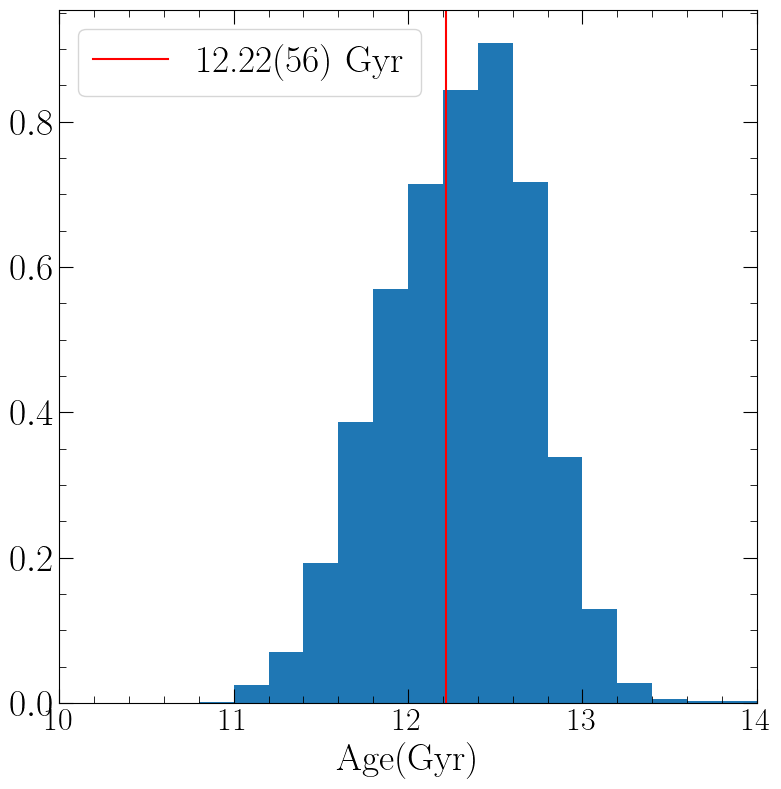}
    \caption{The distribution of age for M55}
    \label{fig:M55_age}
\end{figure}

Fig.~\ref{fig:M55_age} shows the weighted distribution of isochrones marginalized by age using the 2D KS test method. We determine the absolute age of M55 to be $ 12.22 \pm 0.57$ Gyr. The same weighted distribution of isochrones can be marginalized to any parameters. As a result, we also find the distance, reddening, and [Fe/H] for M55. The results for all GCs studied in this paper are summarized in Table~\ref{tab:results}. Table~\ref{tab:results} also shows the consistency in absolute age, [Fe/H], distance, and reddening for each GC despite using three different methods analyzing two different data types. Even though we consider the uncertainties in the theoretical models and observations, due to our carefully designed statistical methods, which leverage more information from the observation data for effective model selection, we can achieve the same level of accuracy as relative age estimation. 

\begin{figure}[!htbp]
    \centering
    \includegraphics[width=0.45\textwidth]{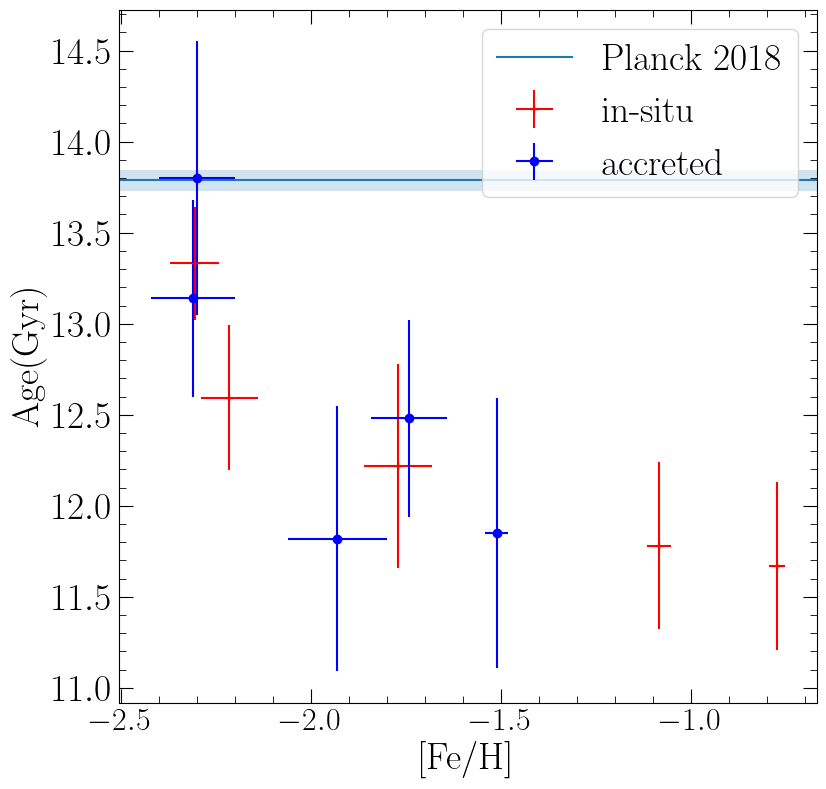}
    \caption{Absolute age vs metallicity plot for 10 Milky Way GCs. The age of the universe, as measured by \citep{planckcollaborationPlanck2018Results2020}, is also plotted with a $3\sigma$ uncertainty.}
    \label{fig:age_metalicity}
\end{figure}

We combine the results from this study with previous studies \citep{yingAbsoluteAgeM922023,yingAbsoluteAgeNGC2024}, and plot the first absolute age-metallicity plot for Milky Way GCs as shown in Fig.~\ref{fig:age_metalicity}. The ages of all $10$ GCs are within $1\sigma$ of the Planck age of the universe\citep{planckcollaborationPlanck2018Results2020}, although M92 has a slightly higher mean age with significant uncertainty. We adopt the in-situ and accreted Milky Way GCs classification based on total energy and angular momentum from \citep{belokurovSituAccretedMilky2024}. Fig.~\ref{fig:age_metalicity} suggests a negative correlation between absolute age and [Fe/H]. However, in-situ and accreted GCs are not distinguishable in this plot. This agrees with \citet{belokurovSituAccretedMilky2024}, who suggest that the in-situ sequence only has a sharp turnover to lower ages at [Fe/H] $\approx -1$ due to the disk spin-up.

\subsection{Error Budget and Individual Monte Carlo Parameter Contributions}
As suggested in the section \ref{age_estimation}, the advantage of this Monte Carlo approach is the capability of getting the marginal distribution of any MC parameters with the weighted distribution of isochrones and estimating the uncertainty of individual MC parameters, as we show in Table~\ref{tab:results}. It is also essential to understand how the variance in individual MC parameters contributes to uncertainty in the age estimation, as it implies the potential improvement in the accuracy with better constraints in the corresponding MC parameter. \citet{yingAbsoluteAgeNGC2024} proposed a dominance analysis-based method to estimate the relative importance of MC parameters. 

The coefficient of determination $R^2$ is a statistical measure used to assess the goodness of fit of a regression model. \citet{lindemanIntroductionBivariateMultivariate1980} proposed a dominance analysis method based on the measure of the elementary contribution of any given variable  $X_j$ to a given subset model $Y(X_u)$ by the increase in $R^2$ that results from adding that predictive variable to the regression model:
\[
\textup{LMG}_j = \frac{1}{d!} \sum_{\substack{\pi \in \textup{permutation} \\ \textup{of} \{1, ..., d\}}} r^2_{Y,(X_j\vert X_{\pi})},
\]
where $u$ is a subset of all indices ${1, ..., d}$ and $X_u$ represents a subset of input. This method, however, suffers from the curse of dimensionality. Since we have $21$ MC parameters, including distance and reddening, we adopt an approximation method \citep{johnsonHeuristicMethodEstimating2000} that utilizes relative weight measures to transform the correlated inputs into uncorrelated variables. This method significantly reduces the computational cost and is used in this analysis. We calculate the Johnson indices for each $8$ GC being analyzed. Each set of Johnson indices decomposes the $R^2$ by the proportion of the variance in age, which the variance of certain MC parameters can explain. We find that the distance, reddening, [Fe/H], [$\alpha$/Fe], Helium diffusion, and mixing length are the most common sources of uncertainties with the highest Johnson indices across all $8$ GCs, which also aligns with our previous studies. 

\begin{figure*}[]
    \centering
    \includegraphics[width=0.9\textwidth]{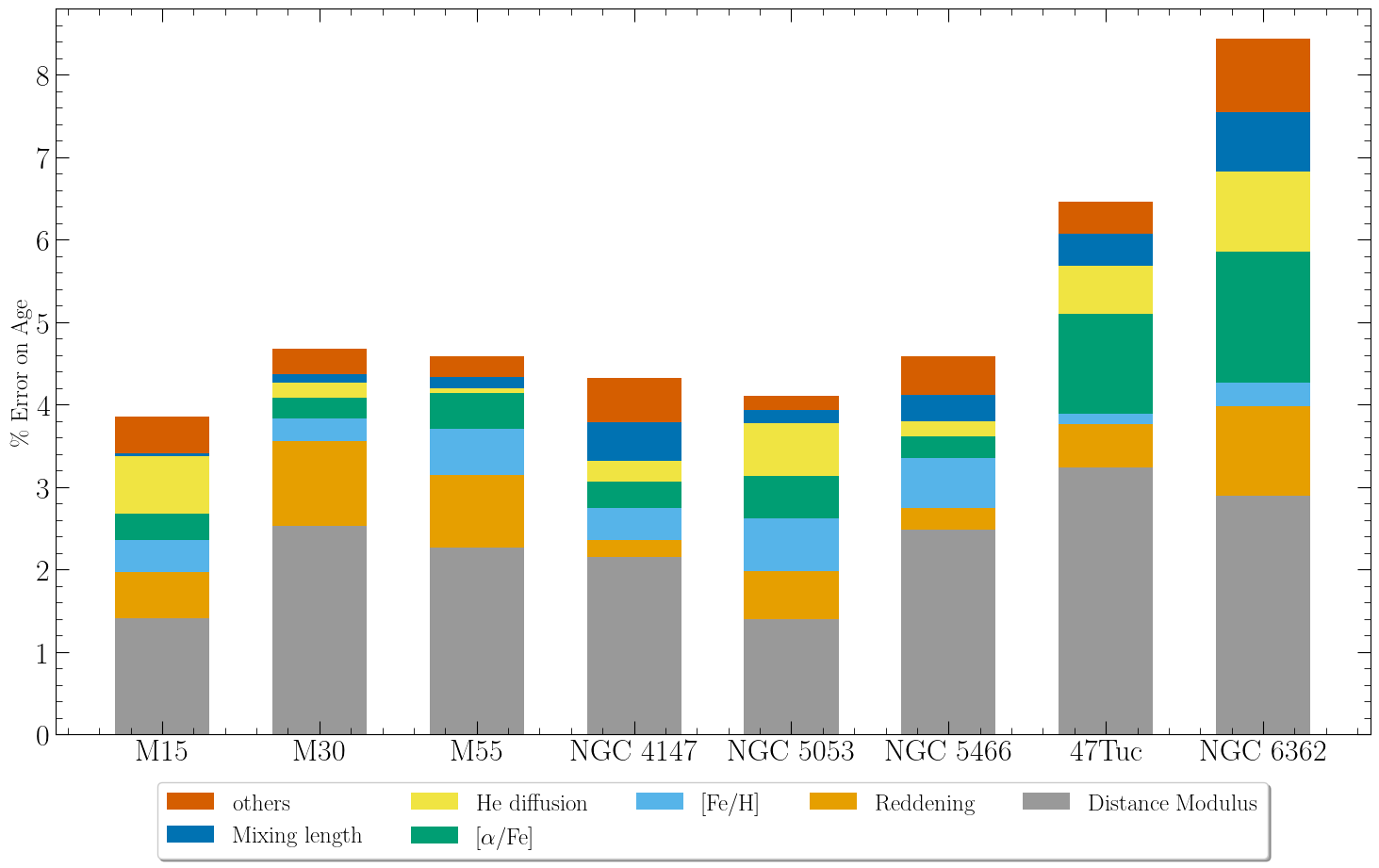}
    \caption{Comparing contributions to the variability of the estimated absolute age of $8$ GCs from each of the Monte Carlo stellar evolution parameters as well as distance modulus and reddening. $\%$ of age error is determined by the Johnson indices multiplied by the coefficient of variance of the absolute age.}
    \label{fig:Johnson}
\end{figure*}

Fig.~\ref{fig:Johnson} illustrates the error budget for the absolute age estimation for all $8$ GCs. The absolute height of each column represents the total $\%$ error in age or the coefficient of variance of the age for the GC. The total  $\%$ error on age is decomposed based on the Johnson indices for the corresponding GC. Others represent the percentage error in age that cannot be explained by any of the eight most important parameters. 47 Tuc and NGC 6362 have a higher percentage of error in age compared to other GCs, with a more significant contribution from $\alpha$ abundance. In this project, we use the library of PHOENIX stellar atmospheres\citep{husserNewExtensiveLibrary2013}, which contains models ranging from [$\alpha$/Fe] $= - 0.2$ to [$\alpha$/Fe] $= 1.2$ with a $0.2$ step size. Therefore, even if [$\alpha$/Fe] is sampled in a continuous space, we have to use the closest PHOENIX model as an approximation. Observations suggest that the mean $\alpha$ abundance for M15, 47 Tuc, and NGC 6362 are between two PHOENIX models (see Table~\ref{tab_indi_MC}). M15 has a strong preference in [$\alpha$/Fe] as the weighted mean of [$\alpha$/Fe] $=0.20$. 47 Tuc and NGC 6362, however, do not show a strong preference in $\alpha$ abundance with almost $50-50$ distribution of [$\alpha$/Fe] in the best-fitting models. The strong correlation between age and $\alpha$ abundance in the best-fitting models is more pronounced in 47 Tuc and NGC 6362.

Fig.~\ref{fig:Johnson} also suggests that distance and reddening are the most dominant sources of uncertainties (contributing more than $50\%$ of the total error on age) for all GCs. This is expected as, unlike other MC parameters, which will have subtle changes in the morphology of the isochrone used to generate the sCMD, distance, and reddening will have a more direct impact as they essentially parallel transport the entire sCMD on the 2D plane. This also aligns with the results of previous studies, which show that distance and reddening strongly correlate with the estimated age. 

\subsection{Constraining Monte Carlo Parameters}

\begin{figure}[!htbp]
    \centering
    \includegraphics[width=0.45\textwidth]{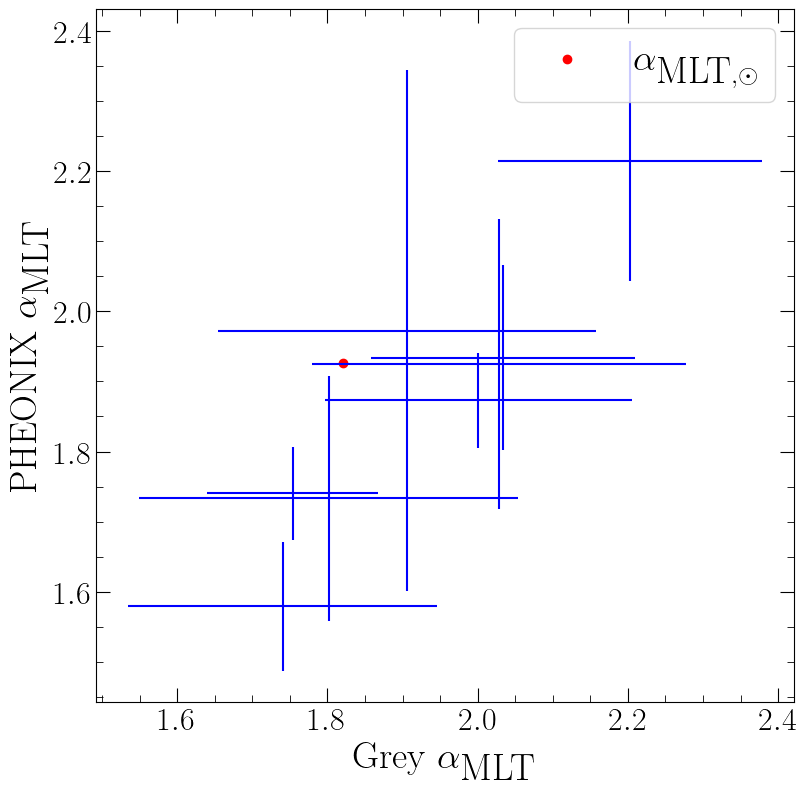}
    \caption{The relation between best-fit $\alpha_{\textup{MLT}}$ for two different surface boundary conditions for 8 GCs (in blue). The red dot is the solar-calibrated mixing length for DSEP\citep{joyceNotAllStars2018}. }
    \label{fig:mlt}
\end{figure}

\begin{figure*}[]
    \centering
    \includegraphics[width=0.9\textwidth]{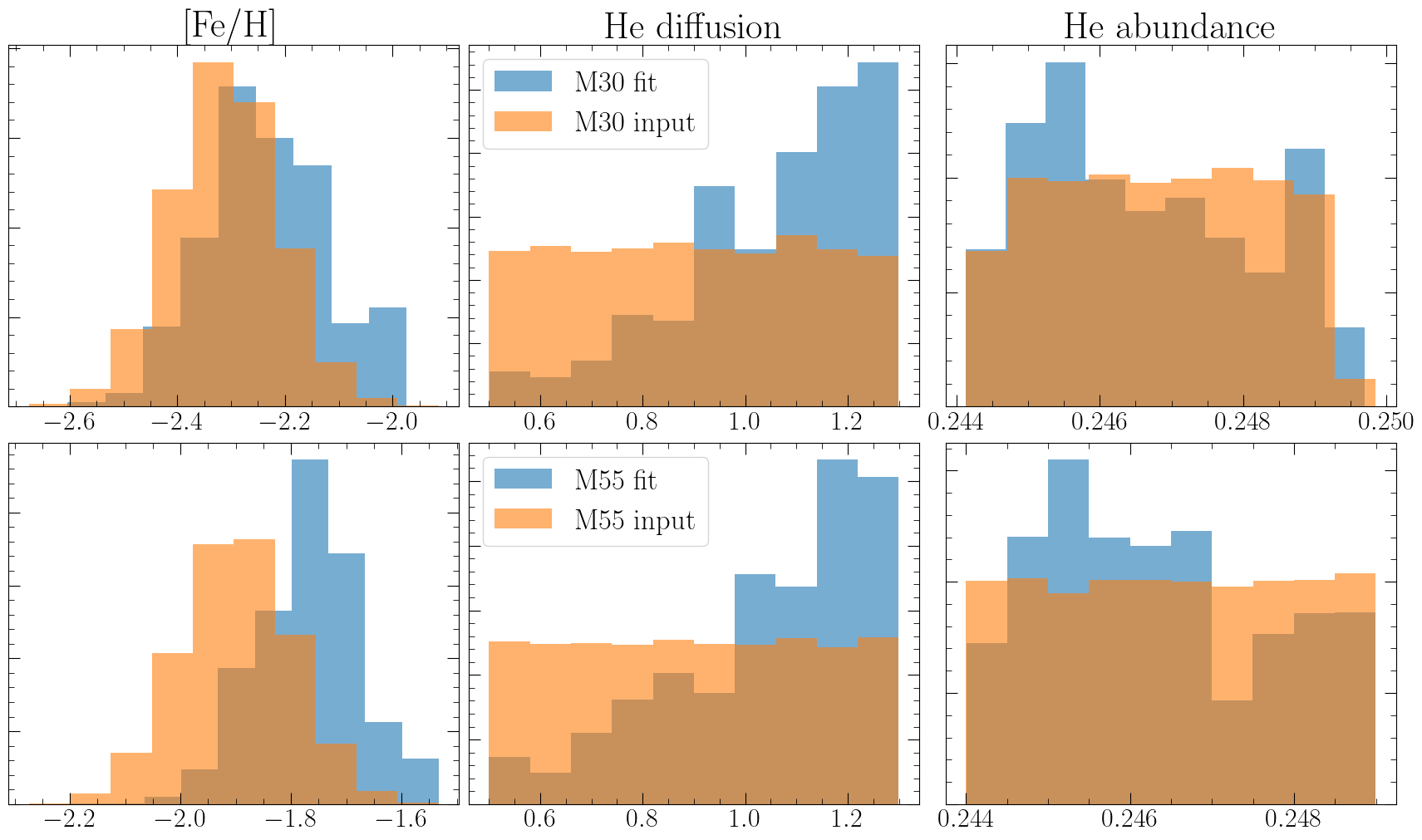}
    \caption{Distribution of [Fe/H], helium diffusion coefficient, and helium abundance for M30 (top panels) and M55 (bottom panels). Orange histograms represent the input distribution, and blue histograms show the weighted distribution of corresponding MC parameters from best-fit models. }
    \label{fig:input_output}
\end{figure*}

One of the advantages of the Monte Carlo approach is the capability to marginalize the posterior distribution, as Table~\ref{tab:results} has already shown $4$ of them. \citep{yingAbsoluteAgeM922023} shows a strong correlation between the surface boundary condition and the mixing length. The exact correlation is found in all $8$ GCs we study in this paper. Moreover, Fig.~\ref{fig:mlt} compares the best-fit mixing length parameter $\alpha_{\textup{MLT}}$ for two different surface boundary conditions: Eddington-Grey model \citep{eddingtonInternalConstitutionStars1926} and PHOENIX model \citep{hauschildtNextGenModelAtmosphere1999, husserNewExtensiveLibrary2013}, and suggests a positive correlation. This is expected as $\alpha_{\textup{MLT}}$ controls convective efficiency despite the choice of boundary condition. Unfortunately, the mixing length parameter does not exhibit an obvious correlation with any other MC parameters, except for boundary conditions, as calibration of $\alpha_{\textup{MLT}}$ remains an empirical task \citep{joyceReviewMixingLength2023}. 

We apply the Kolmogorov–Smirnov 2-sample test for the input distribution and weighted best-fit distribution for each MC parameter for every GC to quantify the capacity to impose constraints. We find 2 GCs with a strong preference for helium diffusion coefficient: M30 and M55, as shown in Fig.~\ref{fig:input_output}. More interestingly, both GCs prefer a small initial helium abundance. The inclination for more metal-rich models offsets the result of lower helium abundance and more efficient helium diffusion compared to observations. This correlation is also observed in a previous study for M92 \citep{yingAbsoluteAgeM922023} in the opposite direction, where metal-poor models are preferred with less effective helium diffusion. \citet{thoulElementDiffusionSolar1994} suggests a $15\%$ uncertainty in helium diffusion from calculating the Coulomb collision integrals. \citet{proffittGravitationalSettlingSolar1991} suggest studying all relevant elements in the solar interior and finding differences in helium diffusion of less than $15\%$. \citet{vinyolesNewGenerationStandard2017} also adopts a $15\%$ uncertainty in helium diffusion for the standard solar model. Therefore, we anticipate that our models will remain physically consistent only if the deviation in helium diffusion from the standard value is less than $30\%$.

\section{Conclusion}
This work determines the absolute ages of $8$ Milky Way globular clusters (GCs). We generate $10,000$ sets of stellar evolution models for each cluster using the state-of-the-art Dartmouth Stellar Evolution Program, sampling stellar evolution parameters via Monte Carlo methods across a $21$-dimensional parameter space. From these models, we construct isochrones spanning ages from $8$ Gyr to $16$ Gyr in increments of $0.2$ Gyr.

For metal-poor GCs, each set of isochrones is calibrated using two main-sequence stars with accurately measured colors and absolute magnitudes obtained from HST ACS photometry and \textit{Gaia} EDR3 parallaxes. We then generate synthetic color-magnitude diagrams (sCMDs) with 4 million data points per isochrone, incorporating photometric uncertainties based on artificial star tests, as well as binary star fractions and the present-day mass function.

We employ the Voronoi binning method and the 2D Kolmogorov–Smirnov method, two statistical measures based on full CMD-fitting techniques, to compare the theoretical models with observational data from the HST ACS GC survey in the F606W and F814W bands. Using uniform priors for distance and reddening, we apply Gaussian process minimization to derive the best-fit distance and reddening for each isochrone, with boundaries carefully selected from literature values. For GCs host detach eclipsing binaries, we use the same set of isochrones to fit the mass, luminosity, and radius derived as an independent test. Despite employing different data types, the methods generally converge on similar conclusions.

Bootstrap resampling is used to estimate the empirical distribution of the test statistics, allowing us to assign weights to each isochrone. By marginalizing over the weighted distribution, we estimate the absolute age, distance, reddening, and other parameters for each GC. 

Our analysis highlights a consistent pattern across all clusters, identifying distance and reddening as dominating sources of uncertainty, collectively contributing over $50\%$ to the total uncertainty in age determination. Additionally, metallicity, alpha abundance, mixing length, and helium diffusion significantly affect the age estimates, underscoring their importance in stellar evolution modeling.

Importantly, we demonstrate a clear trend toward older ages at lower metallicities. We find that the existence of multiple stellar populations in GCs has limited influence on the absolute age of GCs. We suggest that the next generation of stellar evolution models with improved performance and complete cataloging in UV bands, with AS tests that address the PSF modeling issues, is necessary for precisely modeling multiple populations in GCs.

We present the first absolute age-metallicity relation for Milky Way GCs, derived from our expanded sample with absolute ages ranging approximately from $11.5$ to $13.5$ Gyr and typical uncertainties between $0.5$ and $0.75$ Gyr. This relation not only enhances our understanding of GC formation timescales but also provides critical insights into the early accretion history of the Milky Way. Finally, we recommend conducting similar absolute age analyses on metal-rich Milky Way GCs to extend the age-metallicity relation and improve our understanding of galactic formation and evolution.

\section*{Data Availability}
All the mission data used in this paper can be found in MAST: \dataset[https://doi.org/10.17909/t9tg65]{https://doi.org/10.17909/t9tg65} \citep{MAST_ACS}.
All Monte Carlo isochrones constructed for this study are available on Zenodo under an open-source Creative Commons Attribution license: \dataset[doi:10.5281/zenodo.13998637]{https://doi.org/10.5281/zenodo.13998637}

\section*{Acknowledgments}
The authors would like to thank the anonymous referee for the helpful comments and suggestions that significantly improved the paper. This material is based upon work supported by the National Science Foundation under Award No. 2007174, by NASA through AR 17043 from the Space Telescope Science Institute (STScI), which is operated by AURA, Inc., under NASA contract NAS5-26555. MBK acknowledges support from NSF CAREER award AST-1752913, NSF grants AST-1910346, AST-2108962, and AST-2408247; NASA grant 80NSSC22K0827; HST-GO-16686, HST-AR-17028, and HST-AR-17043 from STScI; and from the Samuel T. and Fern Yanagisawa Regents Professorship in Astronomy at UT Austin.

\bibliography{reference}
\bibliographystyle{aasjournal}
\end{document}